\documentclass[a4paper,fleqn,usenatbib]{mnras}
\pdfoutput=1
\usepackage{mathptmx}
\usepackage[T1]{fontenc}
\usepackage{ae,aecompl}
\usepackage{natbib}

\usepackage{graphicx}	% Including figure files
\usepackage{amsmath}	% Advanced maths commands
\usepackage{amssymb}	% Extra maths symbols
\usepackage{epsfig}
\usepackage{multirow}
\usepackage{caption}
\usepackage{subcaption}
\newcommand\msun{\, \rm M_\odot}

\newcommand\myr{{\, \rm Myr}}
\newcommand\gyr{{\, \rm Gyr}}

\newcommand\change{}
\newcommand\changetwo{}

\title[Merging to explain multiple populations and rotation in GCs]{A critical look at the merger scenario to explain multiple populations and rotation in iron-complex globular clusters}

\author[E. Gavagnin et al.]  {
Elena Gavagnin$^{1}$\thanks{E-mail: gavagnin@physik.uzh.ch}, 
Michela Mapelli$^{2}$
and George Lake$^{1}$ \\
$^{1}$Institute for Computational Science, Centre for Theoretical Astrophysics and Cosmology, Universit\"at Z\"urich, Winterthurerstrasse 190, \\ CH-8057 Z\"urich, Switzerland \\
 $^{2}$INAF-Osservatorio Astronomico di Padova, Vicolo dell'Osservatorio 5, I--35122, Padova, Italy\\}

\date{}

\pubyear{2016}

\begin{document}
\label{firstpage}
\pagerange{\pageref{firstpage}--\pageref{lastpage}}
\maketitle
\bibliographystyle{mnras}

% Abstract of the paper
\begin{abstract}

Merging has been proposed to explain multiple populations in globular clusters (GCs) where there is a spread in iron abundance (hereafter, iron-complex GCs). By means of N-body simulations, we investigate if merging is consistent with the observations of sub-populations and rotation in iron-complex GCs. The key parameters are the initial mass and density ratios of the progenitors. When densities are similar, the more massive progenitor {\changetwo dominates the central part of the merger remnant} and the less massive progenitor forms an extended rotating population. The low-mass progenitor can become the majority population in the central regions of the merger remnant only if its initial density is higher by roughly the mass ratio. To match the radial distribution of multiple populations in two iron-complex GCs ($\omega$~Cen and NGC~1851), the less massive progenitor needs to be four times as dense as the larger one. {\change Our merger remnants show solid-body rotation in the inner parts}, becoming differential in the outer parts. Rotation velocity $V$ and ellipticity $\epsilon$ are in agreement with models for oblate rotators with isotropic dispersion.  We discuss several kinematic signatures of a merger with a denser lower mass progenitor that can be tested with future observations. 
\end{abstract}


%%%%%%%%%%%%%%%%% BODY OF PAPER %%%%%%%%%%%%%%%%%%

\section{Introduction}
Over a quarter of the objects in Messier's catalog are 
globular clusters (GCs), yet we still do not know how they were formed.
For many decades GCs were described as stellar systems with homogeneous chemical composition and no age spread, despite early data showing multiple populations in M22 and $\omega{}$ Cen (\citealt{geyer67,cannst73,Harris74,freerodg75,Hesser80}). 

{\it Hubble Space Telescope} data show  a clear bifurcation of colour in the main sequence (MS) of $\omega{}$ Cen \citep{anderson97}, with more recent data showing at least four distinct red giant branches (RGBs, \citealt{lee99, pancino00}). 
Currently, most observed GCs show signatures of multiple populations, both in the Milky Way \citep{Gratton04, Carretta07,Kayser08,Anderson09,Carretta09b,Carretta10,Carretta15,Pancino10,Milone10,Milone12,Milone13} and in the Magellanic Clouds \citep{Milone08}. 

Most GCs contain stars with similar heavy-element abundances (especially [Fe/H]), but large ($>0.5$ dex) star-to-star abundance variations for elements lighter than Si (e.g. \citealt{cohen78,peterson80,sneden91,kraft92,Gratton01,Carretta09b,johnson15}). Moreover, the variations of light-element abundances are anti-correlated with one another (e.g. the O$-$Na anti-correlation, \citealt{Gratton01}). This phenomenology is generally considered {\change to be due to} internal enrichment by proton capture H-burning reactions at high temperature (e.g. \citealt{Gratton04}).

{\change A minority of GCs also show} significant  Fe abundance variations. In particular, $\omega{}$ Cen \citep{norris95,lee99,bellini10,dorazi11,pancino11}, M22 \citep{hesser77,marino09,Lee15}, M2 \citep{piotto12,lardo13,milone15}, M54 \citep{sarajedini95,bellazzini08,carretta10b}, NGC 1851 \citep{yong08,milone09,carretta10c,Carretta11}, NGC 5286 \citep{nataf13,marino15}, NGC 5824 \citep{saviane12,dacosta14}, Terzan 5 \citep{ferraro09,massari14} and M19 \citep{johnson15} are labelled as 'iron-complex' GCs, because they have (i) a spread\footnote{Recent studies highlight the possibility that the  [Fe/H] spread is spurious, at least in some GCs (e.g. M22, \citealt{mucciarelli15}), because spectroscopically derived Fe abundances might be inaccurate due to non-local thermodynamical equilibrium effects.} in [Fe/H] exceeding $\sim{}0.10$ dex,
(ii) multiple photometric sequences, and (iii) a significant abundance spread for both light and heavy elements \citep{johnson15}. Iron-complex GCs {\change differ from other GCs} in several ways.  In most GCs, the stellar population showing no enrichment by proton capture accounts for about one third of the total GC mass, with little spread among GCs \citep{carretta09c}. In contrast, in the iron-complex GCs, {\change the ratio between the metal-poor and the} metal-rich population changes from cluster to cluster. For example, in M19 the metal-poor component is $\sim{}50$ \% of the entire population \citep{johnson15}, whereas $\sim{}96$ \% of spectroscopically studied stars in M2 belong to the metal-poor component \citep{milone15}.  Moreover, in the vast majority of GCs, the proton-capture enriched population is more radially concentrated than the most numerous one. In the iron-complex GCs the metal-poor population can be either more concentrated ($\omega$ Cen, \citealt{Bellini09}) or less concentrated (NGC~1851, \citealt{Carretta11}) than the metal-rich one.

Several theoretical models have been proposed to explain the multiple populations (\citealt{Bastian13}; see \citealt{Renzini08} for a review).  A first class of models appeals to multiple star-formation events. {\change After first population stars form out of pristine, metal-poor gas,} the second population of stars might form from either the ejecta of asymptotic giant branch (AGB) stars \citep{D'Ercole08} or fast rotating massive stars (FRMS, \citealt{Decressin07}). In the `AGB scenario', winds and supernovae (SNe) of the first population evacuate the residual gas. After $\sim 30 \myr$, low-velocity winds from AGBs enriched in He and s-process elements start accumulating at the centre and form the second {\change population}. However, the predicted mass of the second population is an order of magnitude lower than what is observed, requiring a top-heavy first {\change population} initial mass function and an unusually efficient second {\change population} star formation. 

A second model, called the `early disc accretion model' \citep{Bastian13}, proposes that the two populations formed during the same star formation episode, but underwent different chemical enrichment. This model requires very fast mass segregation and gas evaporation. With rapid mass segregation, the most massive stars sink to the centre where high-mass stars in interacting binaries eject the primary's He-enriched envelope. This material pollutes the circumstellar discs of low mass stars that are still accreting, so they will grow in mass  thanks to these ejecta from more massive (but still same-generation) stars.
The main drawbacks of this model are disc lifetime and uniformity of enrichment.  Even if circumstellar discs survive for the required $5-10 \myr$ \citep{DeMarchi13}, the ``rainfall" of enriched material onto them is unlikely to be uniform \citep{Kruijssen14}. 

 All the aforementioned scenarios are aimed to explain multiple populations with no or negligible iron spread, while they fail to reproduce the [Fe/H] variations observed in the iron-complex GCs. So far, the only proposed scenario that can naturally account for a metallicity spread is the merger between GCs \citep{SugimotoMakino89, Makino91,VanDenBergh96,AmaroSeoane13,Pasquato16}. In this scenario, the different metallicities are signatures of the progenitors and can be used as a tag to make predictions about the distribution and dynamics within the final merger remnant. Iron abundance is, in this respect, a good tag to identify uniquely the different populations.

The merger scenario  might be consistent with the oldest metal-rich stars in $\omega{}$ Cen being a few Gyrs older than the oldest metal-poor stars \citep{Villanova14}, a circumstance that is against the predictions of self-enrichment scenarios. Furthermore, a merger can explain the kinematical differences in the velocity dispersion of the calcium-weak and calcium-strong RGB stars in M22 \citep{Lee15}. The merger scenario has been proposed also for NGC~1851, where the most metal-rich population is less concentrated than the metal-poor one \citep{VanDenBergh96, carretta10c,Carretta11,Bekki12}. 

{\change Another advantage of the merger scenario is that it can account for signatures of rotation in GCs, which have been observed in several GCs with multiple populations, both with (e.g. $\omega$ Cen, M2, M22, M54, NGC1851, \citealt{pancino07,Lee15,pryor86,kimming15,Bellazzini12,Lardo15}) and without (e.g. \citealt{Fabricius14}) {\changetwo a} metallicity spread.} If the two progenitors have non-zero relative orbital angular momentum, the merger remnant will likely preserve a signature of rotation in the merger plane. However, there is no  evidence that GCs with a metallicity spread (the best-candidate merger remnants) have systematically higher rotation than the other GCs. Moreover, other physical mechanisms can account for rotation in GCs (e.g. \citealt{mastrobuono13,vesperini14,bianchini15}).

The main problem for the merger scenario is that  two GCs are expected to merge only if their relative velocity is smaller than (or of the same order of magnitude as) their velocity dispersion.  The largest GC in the galaxy, $\omega{}$ Cen, has a dispersion of $\sim{8}$ km s$^{-1}$, with typical values being $\sim{4-6}$ km s$^{-1}$. The relative velocities of current GCs in the Milky Way halo are at least one order of magnitude larger than these values. This means that a merger between two GCs that are in the halo of our galaxy is extremely unlikely. Two GCs {\change will} have a sufficiently low relative velocity to merge only if they formed in a small dwarf galaxy or in the same molecular cloud. {\change However}, if the two progenitor clusters formed in the same molecular cloud and merged slightly after their formation, it is difficult to explain why the two populations have a different proton-capture enrichment and even a different metallicity.
{\change Therefore, GCs in small dwarf galaxies represent the most likely scenario where GC mergers will produce clusters that have a spread in metallicity.}

We take a critical approach to the merging scenario by examining how the initial mass and density ratios of the progenitors affect  the distribution and concentration of the sub-populations in the remnant (Section \ref{results1}). Moreover, we also examine the rotation signature of the merger product and we show that the profile of rotation is related to the initial density ratio of the progenitors (Section \ref{results2}).  
In the event of equal-mass mergers, we  expect that the denser initial progenitor will be more centrally concentrated in the remnant.  In the case of unequal-mass mergers, the more massive progenitor will be closer to the centre than the less massive progenitor and hence be more concentrated.  We examine how the density ratio can counter the mass ratio.

This paper is organised as follows. In Section \ref{methods}, we describe the numerical tools and the initial conditions adopted. In Section \ref{results}, we present the main results of this work. Section \ref{discussion} is dedicated to the discussion and conclusions.

\section{Methods and simulations}\label{methods}

\begin{table*}
\begin{center}
\caption{Initial conditions of the simulations. Run (column 1): identifying name of the run, `M' stands for mass ratio ($M1/M2$), `$\rho$' for density ratio ($\rho_1$/$\rho_2$), both followed by the values assumed, e.g. M4$\rho$0.25, means mass ratio = 4, density ratio = 0.25; column 2:  progenitor identifier; $N$ (column 3): number of particles; $M$ (column 4): total mass of the progenitor; $R_{\rm V}$ (column 6): virial radius; $D$ (column 7): initial distance between the progenitors' centres of mass; $r_{\rm peri}$ (column 9): orbital pericentre; $V$ (column 10): initial relative velocity.} \leavevmode
\label{table_IC}
\setlength{\tabcolsep}{12pt}
\begin{tabular}[]{l  l  l  l  l  l l  l l l l}
\hline
Run
&
& $N$
& $M$
& $R_{\rm V}$
& $D$
& $r_{\rm peri}$
& $V$\\

&
&
& [$\msun$]
& [pc]
& [pc]
& [pc]
&[km s$^{-1}$]\\
\hline \hline
\noalign{\vspace{0.1cm}}

&GC2 	&20k 		&$10^5$	&4			&			&	   &	\\

\hline \hline
\noalign{\vspace{0.1cm}}

M1$\rho$0.25	&GC1 	&20k			&$10^5$	&6.34		&111.58		&5.17   &5.59\\ 
				  																								
\hline
M1$\rho$0.50 	&GC1 	&20k			&$10^5$	&5.03		&88.5		&4.51   &6.27\\ 
    										
\hline
M1$\rho$1 	&GC1 	&20k			&$10^5$	&4			&70.4		&4   &7.03\\ 
    										
\hline
M1$\rho$2 	&GC1 	&20k			&$10^5$	&3.17		&70.4		&3.59   &7.03\\ 
    										
\hline
M1$\rho$4 	&GC1 	&20k			&$10^5$	&2.52		&70.4		&3.26   &7.03\\ 
    										
\hline
\hline

M2$\rho$0.25	&GC1 	&40k			&$2 \cdot 10^5$		&8	&140.8		&6   &6.09\\ 
				  						
\hline
M2$\rho$0.50 	&GC1 	&40k			&$2 \cdot 10^5$		&6.34	&111.58		&5.17   &6.84\\ 
    										
\hline
M2$\rho$1 	&GC1 	&40k			&$2 \cdot 10^5$		&5.03	&88.5		&4.51   &7.68\\ 
    										
\hline
M2$\rho$2 	&GC1 	&40k			&$2 \cdot 10^5$		&4	&70.4		&4   &8.61\\ 
    										
\hline
M2$\rho$4 	&GC1 	&40k			&$2 \cdot 10^5$		&3.17	&70.4		&3.59   &8.61\\

\hline
\hline

M4$\rho$0.25	&GC1 	&80k			&$4 \cdot 10^5$			&10	&176			&7   &7.03\\ 
				  						
\hline
M4$\rho$0.50 	&GC1 	&80k			&$4 \cdot 10^5$			&8	&140.8		&6   &7.86\\ 
    										
\hline
M4$\rho$1 	&GC1 	&80k			&$4 \cdot 10^5$		&6.34	&111.58		&5.17   &8.83	\\ 
    										
\hline
M4$\rho$2 	&GC1 	&80k			&$4 \cdot 10^5$		&5.03	&88.5		&4.51   &9.92	\\ 
    										
\hline
M4$\rho$4 	&GC1 	&80k		 	&$4 \cdot 10^5$			&4	&70.4		&4   &11.12	\\

\hline
\hline

\end{tabular}

\end{center}
\footnotesize{}
\end{table*}

We used the {\sc starlab} public software environment \citep{Starlab} ported to GPUs \citep{Gaburov09} to run the simulations. To investigate the role of the relative masses and densities of the progenitors, we performed a grid of simulations varying the mass ratio, i.e. $M_1/M_2$ (where $M_1$ is the mass of GC1 and $M_2$ is the mass of GC2) and the density ratio, i.e. $\rho_1/\rho_2$ (where $\rho_1$  and $\rho_2$ are the densities measured at the virial radius of GC1 and  GC2, respectively).

We consider mass ratios of 1, 2, 4, with density ratios of 0.25, 0.5, 1, 2, 4. The range is motivated by the ratio of the populations in GCs {\change \citep{johnson15,milone15}} and the absence of strong correlations between luminosity and density in present-day GCs  \citep{Harris96}. 
The GCs are modelled as non-rotating spherical King profiles \citep{King66} with central dimensionless potential $W_0 = 5$ (this sets the core radius $R_{\rm c} =  0.41\,{}R_{\rm V} $).
The second GC (GC2)  is always composed of 20 000 particles of equal mass $m_{\ast}=5\,{}\msun$ for a total mass of $10^5$ $\msun$. Its virial radius $R_{\rm V} = 4$ pc.
The first GC (GC1) is varied to set the mass ratio and density ratio. {\change To double (or quadruple the mass) of GC1, we double (or quadruple) the number of particles.}

The density ratio is set by adjusting the virial radius of GC1, e.g. in the run $M2\rho{}1$ the GC1 has twice the mass as GC2 and $R_V$ of GC1 is $\sim{}1.26\times{}$ larger than the one of GC2, so that the density ratio between the two clusters is 1.  We {\change note} that, assuming a fixed value for $W_0$, the density ratio is the same at every fiducial radius, i.e. the core radius ($R_{\rm c}$), the tidal radius ($R_{\rm t}$) and the virial radius ($R_{\rm V}$).

To prevent strong encounters and binary formation, we adopt a gravitational softening $\epsilon = 0.1 R_{\rm V} $  of the {\change progenitor with the smallest radius}.  The initial binary fraction is zero and binaries do not form with this softening. We omit stellar and binary evolution to {\change minimise the amount of free parameters in these models}.  
Stellar and binary evolution might affect the structural properties of GCs \citep{Chernoff90,Mapelli13,MapBres13,trani14, Gieles13,Sippel12} and will be considered in a follow up study.  
Stars initially belonging to each of the two progenitors are ``tagged''  with a different metallicity flag.
 Initial conditions (ICs) are summarised in Table \ref{table_IC}.

The two GCs are initially set on a parabolic orbit. 
To define the parabolic orbit we fixed the minimum encounter distance (in the point-mass assumption), i.e. the pericentre $r_{\rm peri}$,  to be half the sum of the virial radii of the two progenitors GC1, GC2 [$r_{\rm peri} = 0.5 \, (R_{V1}+ R_{V2})$]. 
The initial distance $D$  between the progenitors is four times the maximum value between $R_{\rm t,1}$ and  $R_{\rm t,2}$, where $R_{\rm t,1}$ and $R_{\rm t,2}$  indicate the tidal radius of GC1 and GC2, respectively. The initial relative velocity is then calculated as the escape velocity at the initial position. 

We choose a parabolic orbit because it is a representative case for mergers \citep{Alladin65, White78}.  
Hyperbolic encounters (with relative velocity much larger than the GC velocity dispersion) are the most common, as the phase space for encounters increases with the cube of the velocity of encounter and the cube of the impact parameter.  However, the 
probability of merging encounters is sharply truncated (by failure to merge) when the orbits become very weakly hyperbolic.  
In contrast, the two GCs will merge on a shorter timescale if they are on a bound orbit, but bound orbits are associated with smaller values of the velocity. As we mentioned in the introduction, the main drawback of the merger scenario is that the observed relative velocities between GCs are generally larger than the value needed for a merger to be successful. Thus, we consider bound orbits very unlikely. In summary, a parabolic orbit is representative of the most likely orbits leading to a merger.

The half-mass relaxation time is \citep{Spitzer87}
\begin{equation}\label{eq:trlx}
t_{\rm rlx} \sim 3 \times 10^8 {\rm yr} \left(\frac{M}{10^5 \msun}\right)^{1/2} \left(\frac{R_{\rm hm}}{3\,{}{\rm pc}}\right)^{3/2}  \left(\frac{m}{5 \msun}\right)^{-1}  \left(\frac{\ln \Lambda}{3}\right)^{-1},
\end{equation} 
 where $R_{\rm hm}$ is the initial half-mass radius, $M$ the total mass, $m$ is the particle mass and $\ln \Lambda$ is the Coulomb logarithm (set here by the system size and gravitational softening). For our progenitors, the relaxation timescales are between $400\myr$ and $1.7\gyr$. The initial crossing time at the virial radius in the equal mass, equal density progenitors is $\sim$  0.4 $\myr$ and scales as $\rho ^ {-1/2}$.
We run our simulations for 550 Myr.   This is less than one half-mass relaxation timescale characteristic of the merger product in all cases, but two-body encounters have likely contributed to isotropising the velocities in the remnant.

\section{Results}\label{results}

We examine the relative concentration and rotation of the two different populations in the merger remnant.  

\subsection{\change Relative Concentration}\label{results1}
                 
\begin{figure*}
\centering
\includegraphics[width=.3\textwidth]{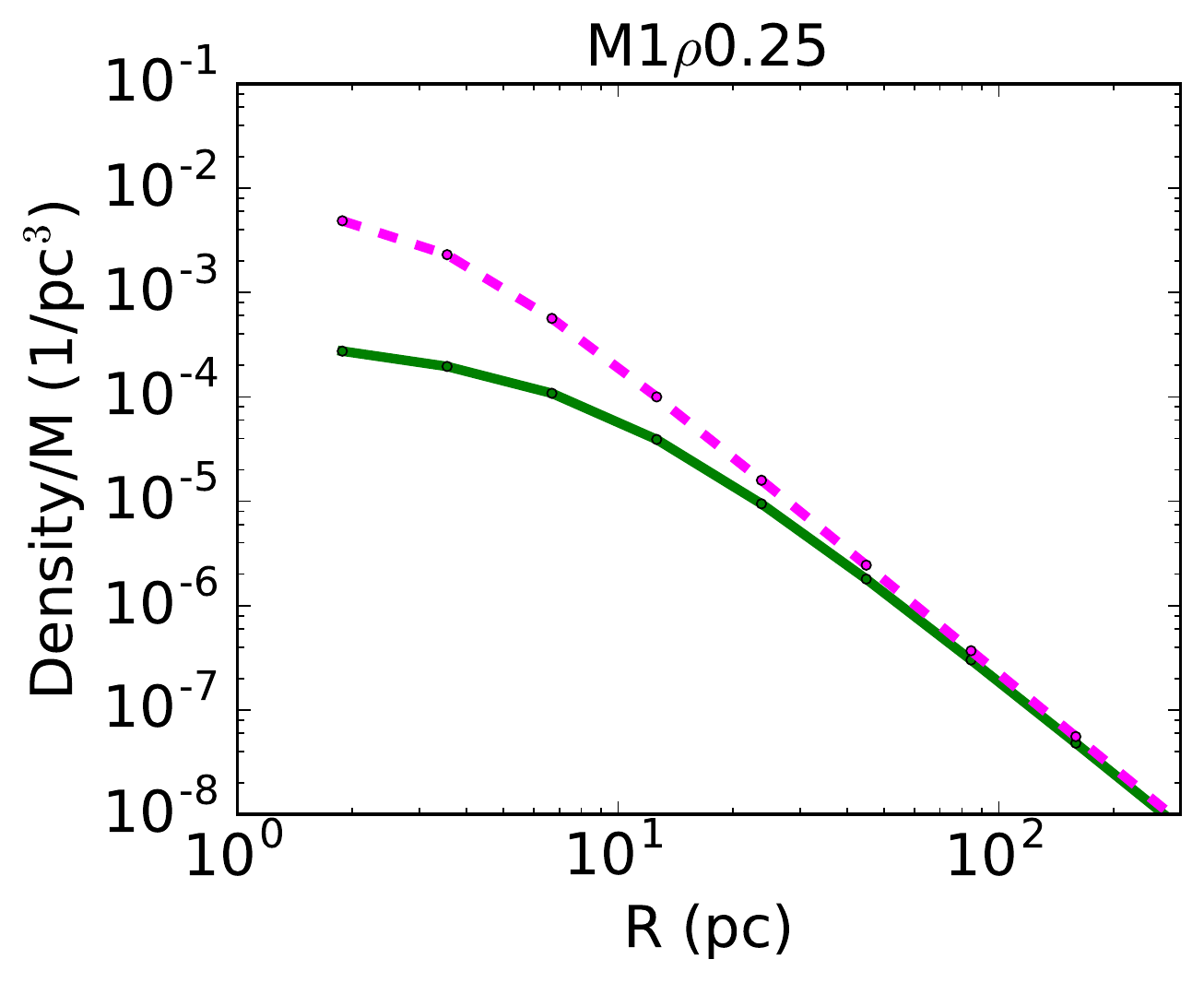}\quad
\includegraphics[width=.3\textwidth]{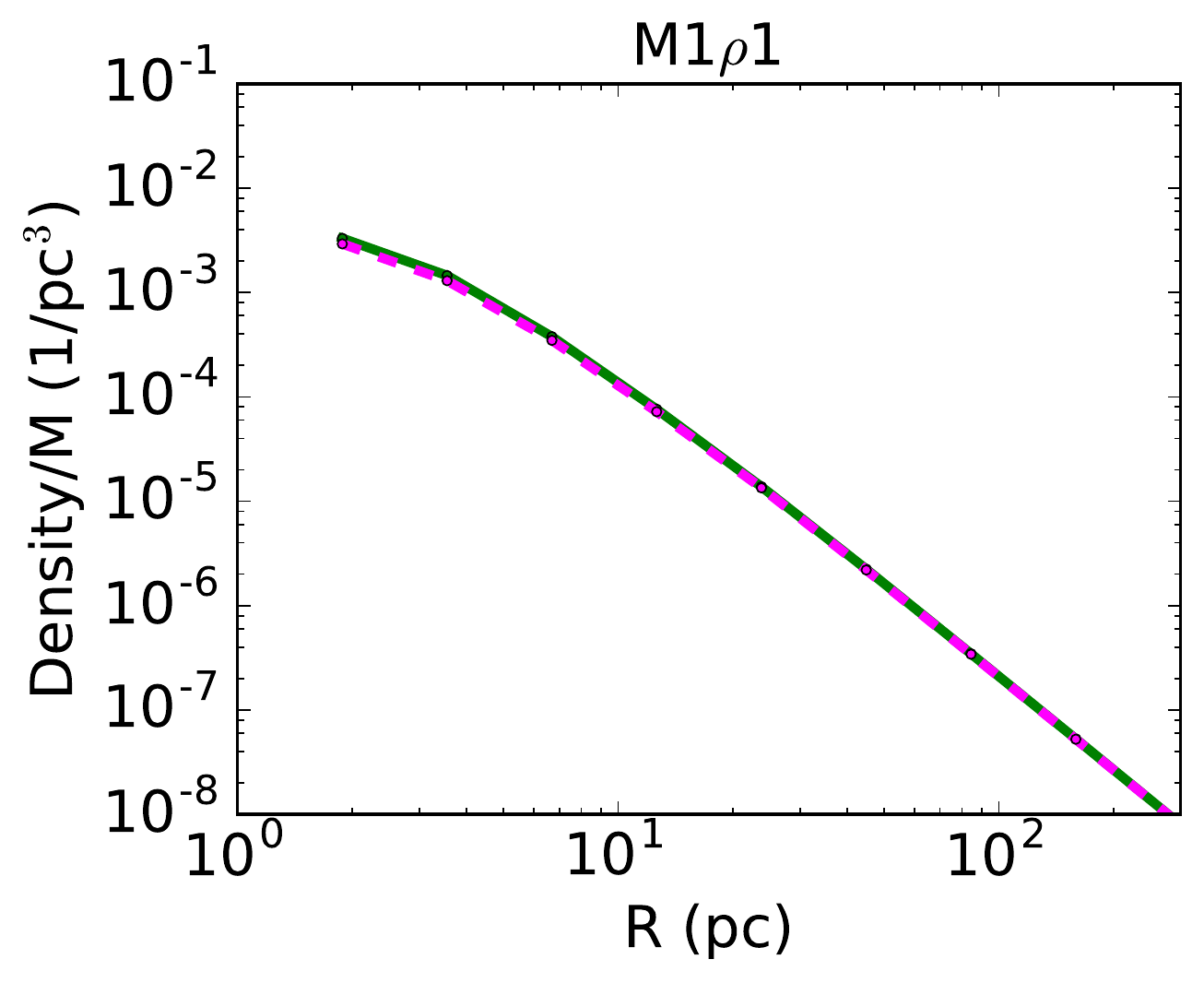}\quad
\includegraphics[width=.3\textwidth]{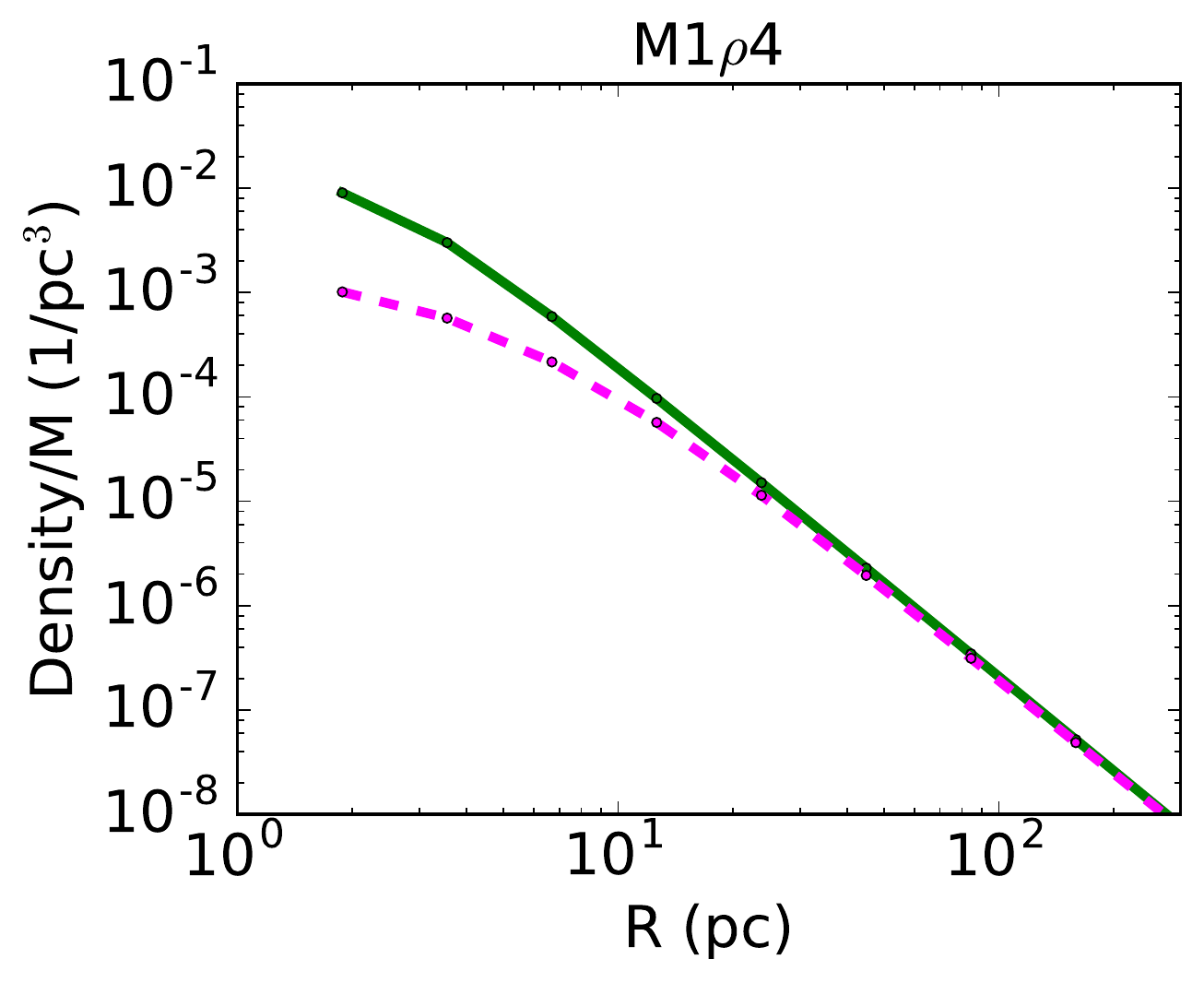}

\medskip

\includegraphics[width=.3\textwidth]{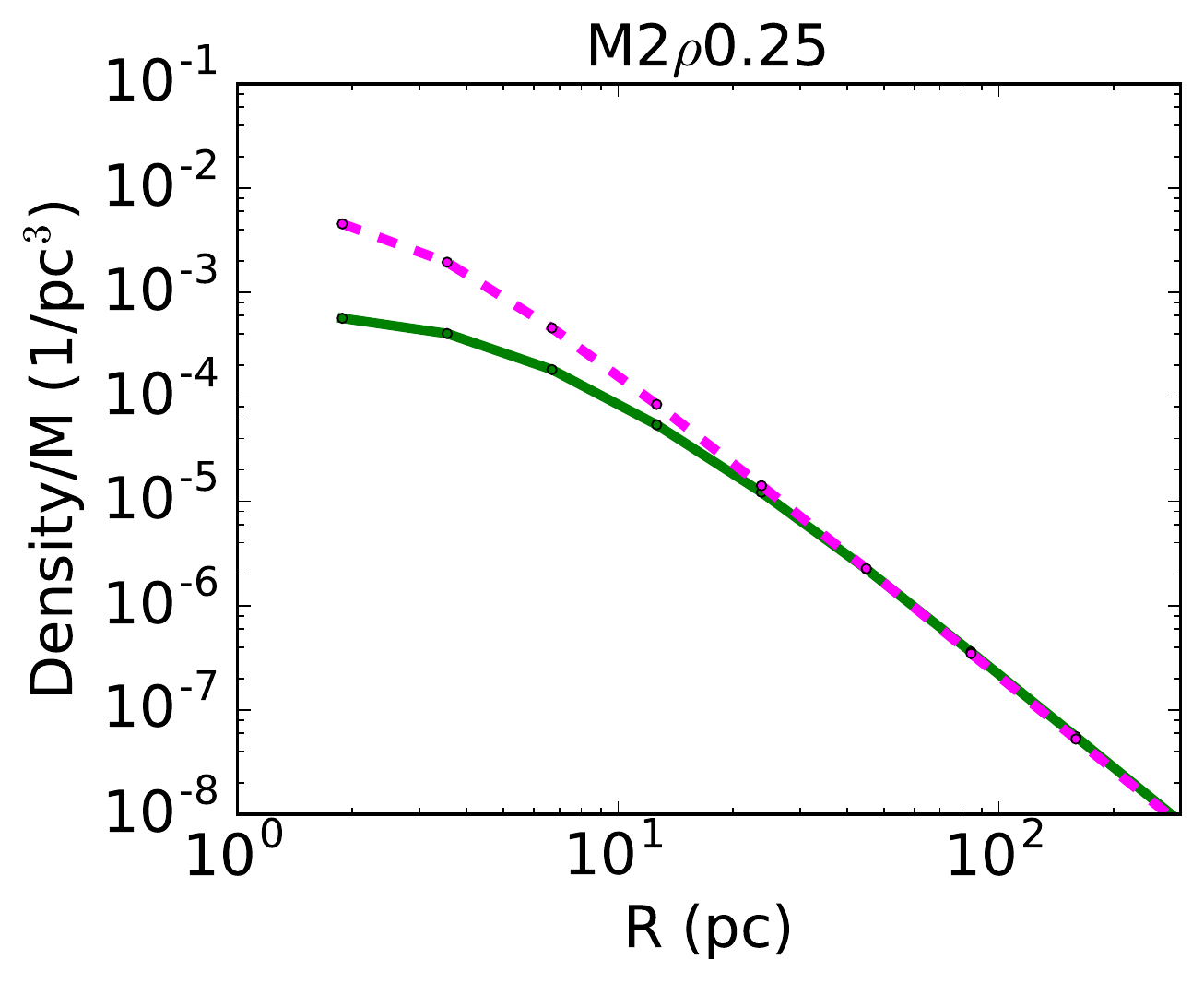}\quad
\includegraphics[width=.3\textwidth]{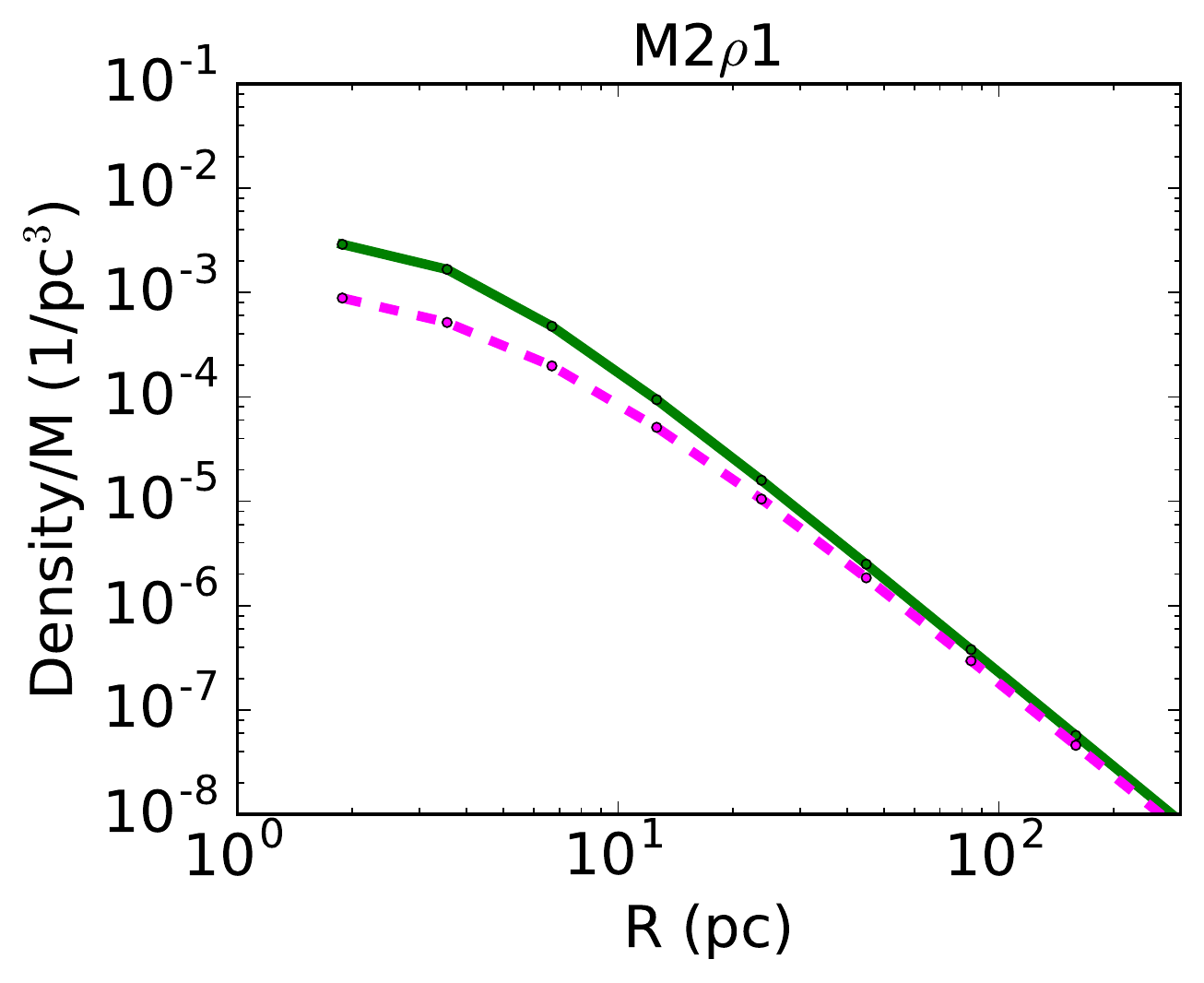}\quad
\includegraphics[width=.3\textwidth]{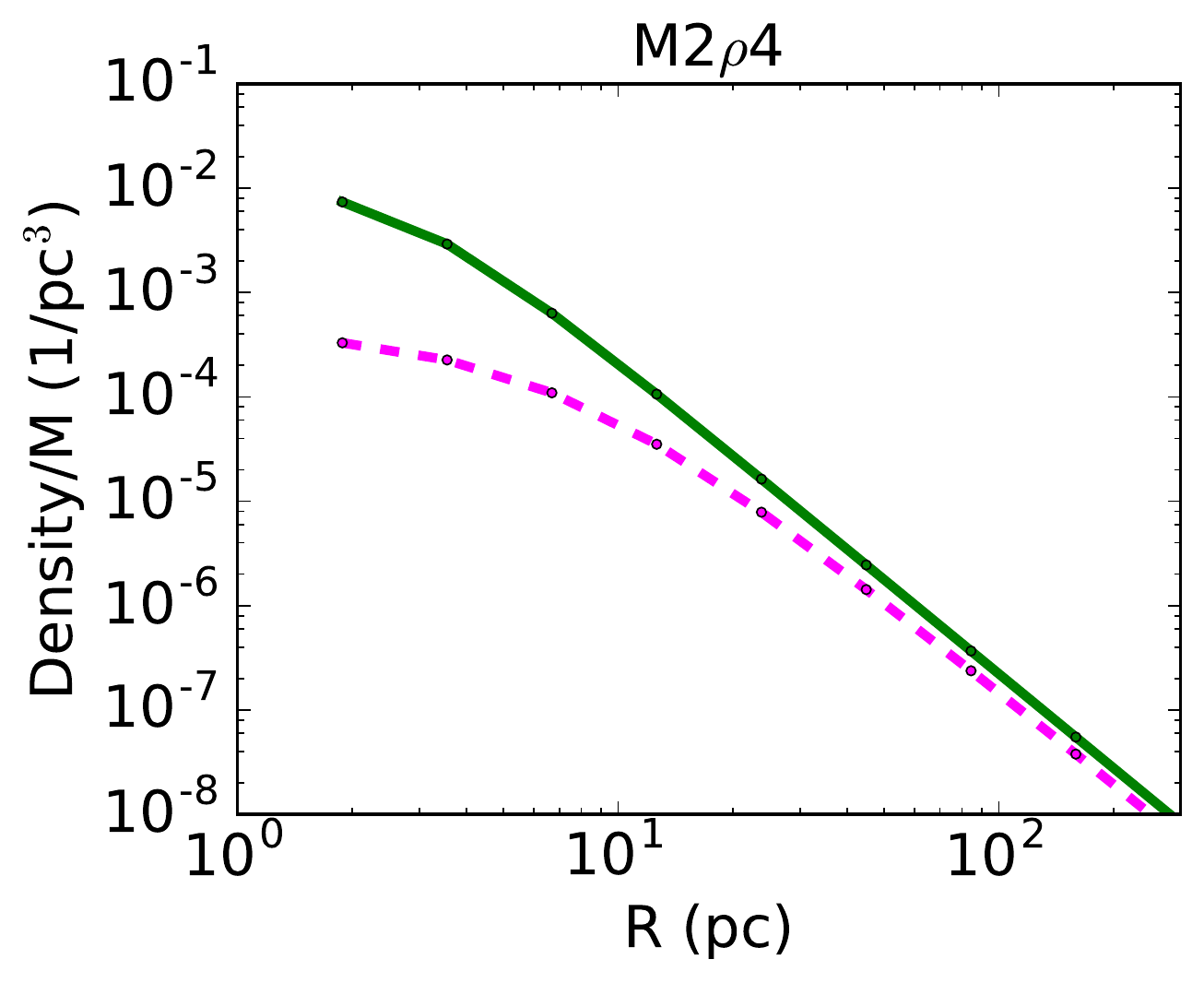}

\medskip
\includegraphics[width=.3\textwidth]{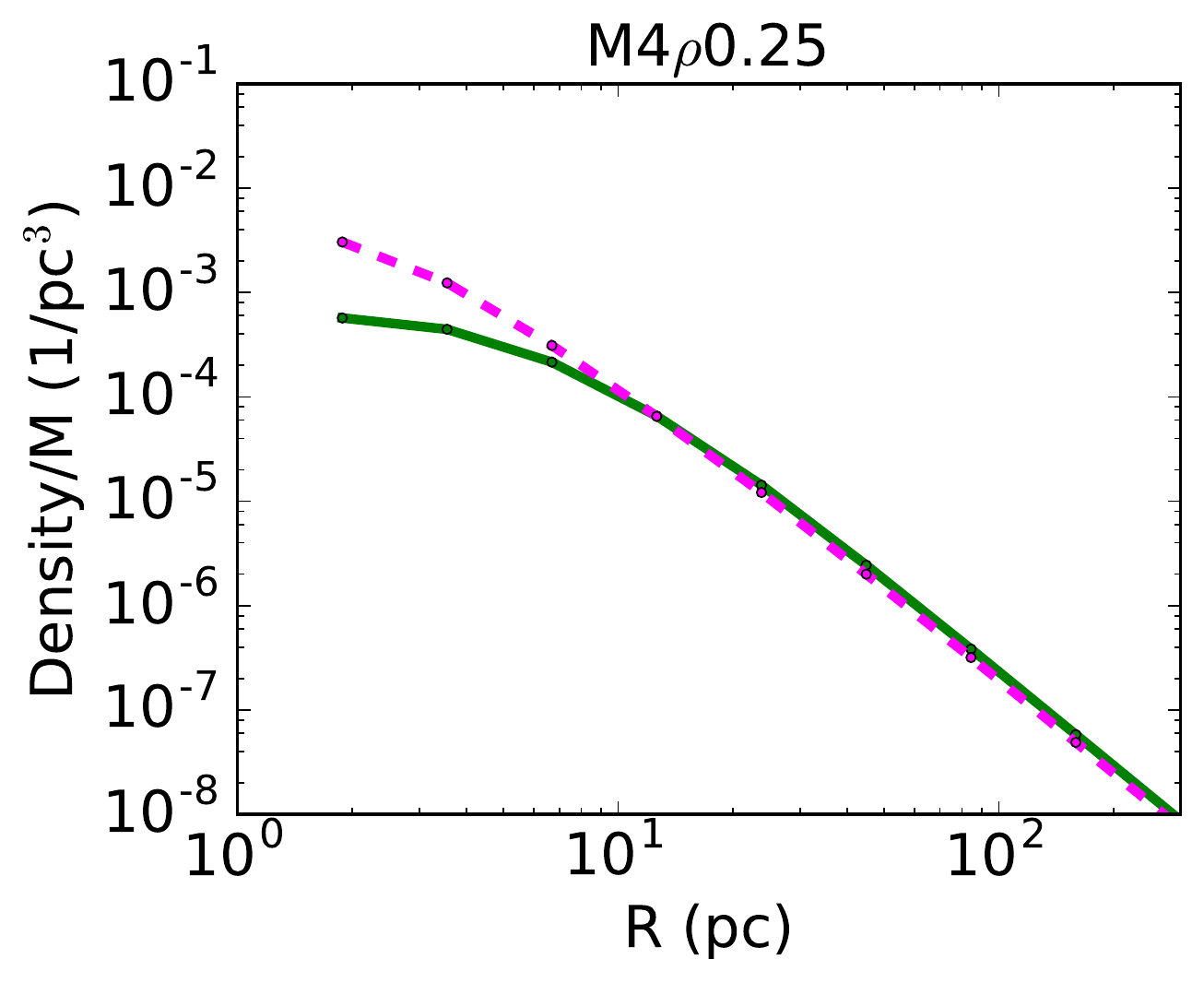}\quad
\includegraphics[width=.3\textwidth]{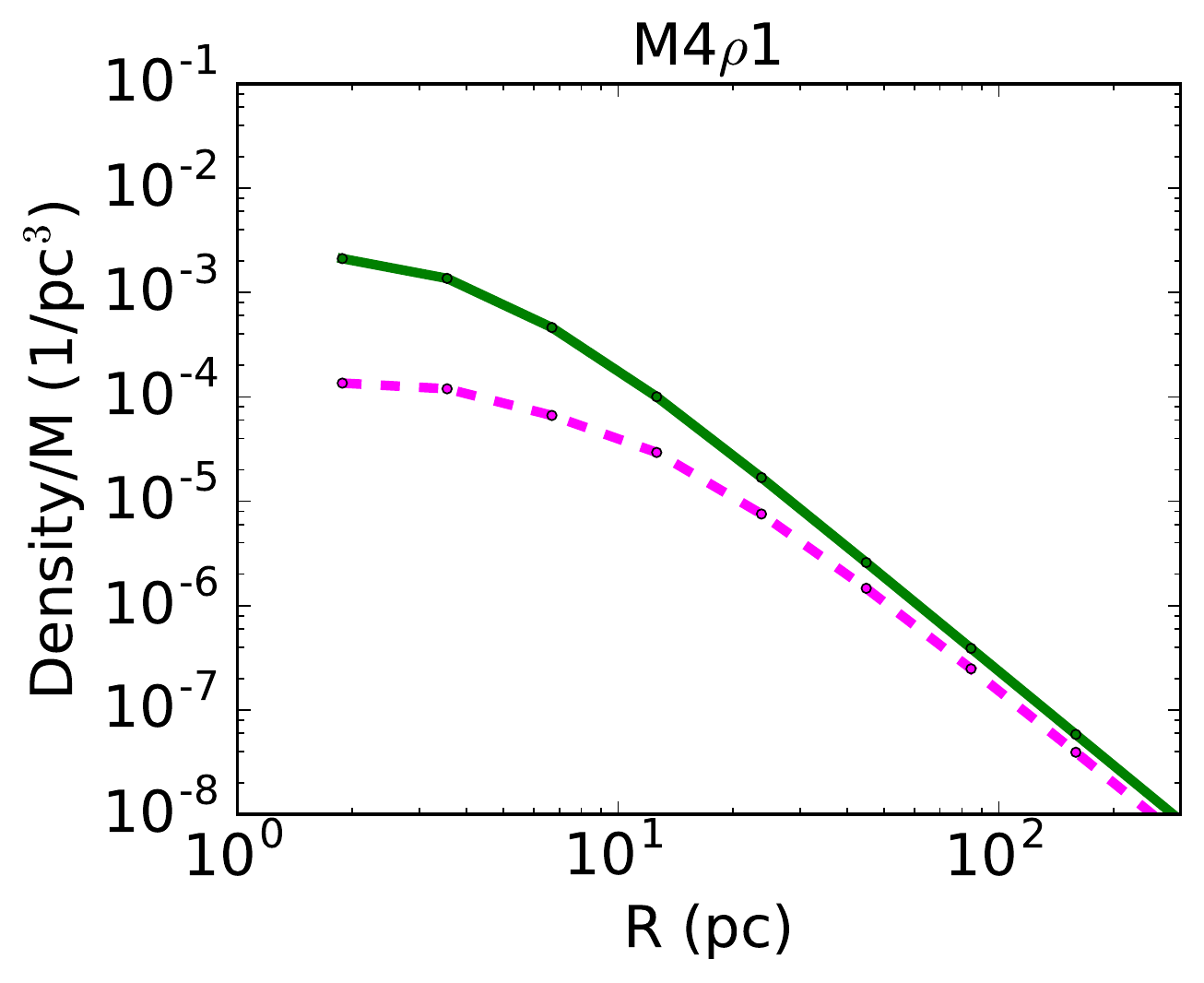}\quad
\includegraphics[width=.3\textwidth]{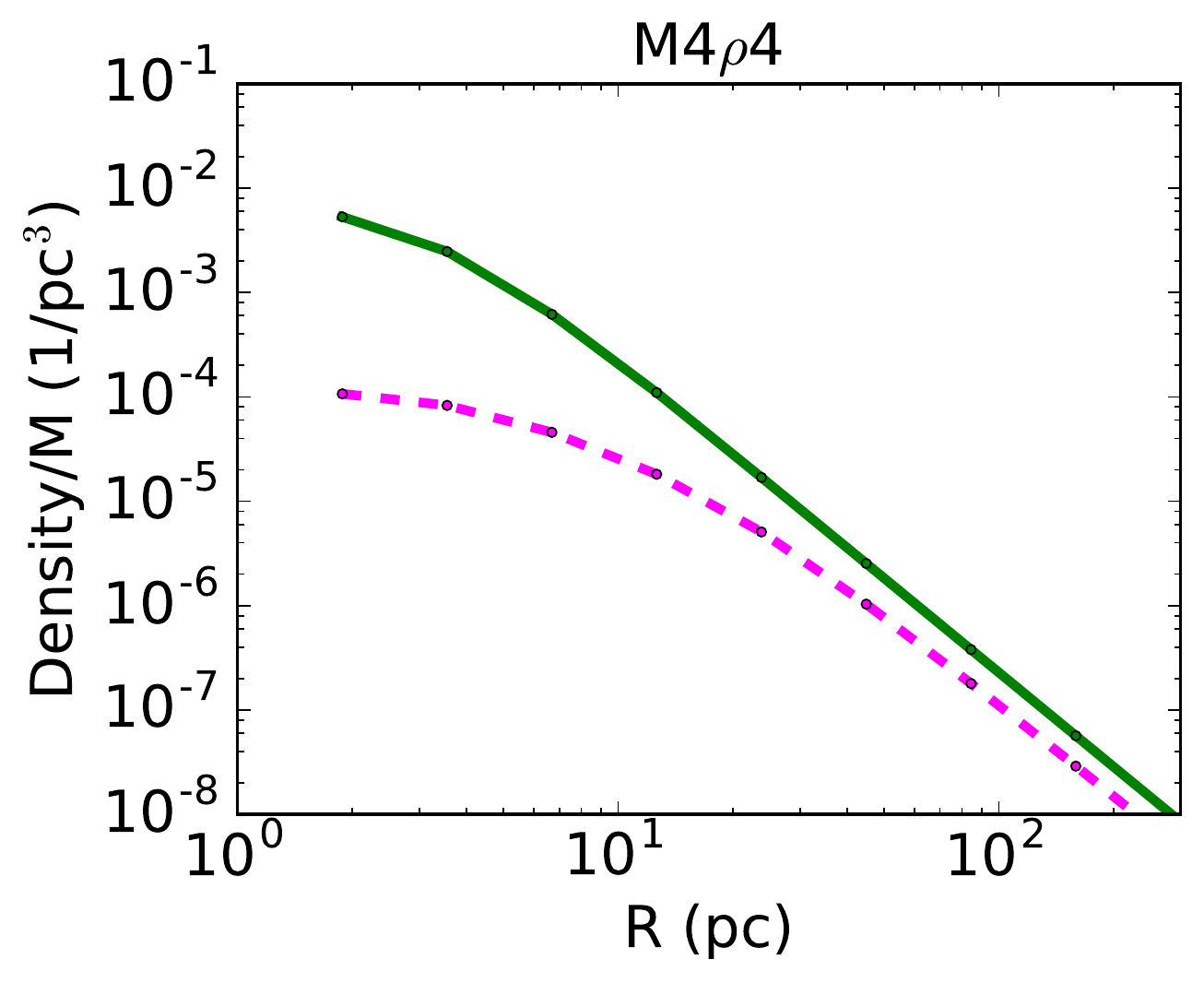}

\caption{Normalised density profiles of the two populations in the final merger remnant. Note that the profiles all look like smooth King models.  Solid green line refers to GC1, dashed magenta line to GC2. Each profile has been normalised by the mass of the associated progenitor.   
The codename on the top of each plot refers to the run considered: `M' stands for mass ratio $M_1/M_2$ followed by its value, `$\rho$' for density ratio $\rho_1$/$\rho_2$ followed by its value. From top to bottom the mass ratio increases by a factor of 2 every row and from left to right the density ratio increases by a factor of 4 every column.  }
\label{density_normal}
\end{figure*}

We plot the relative concentration using  normalised density profiles of the sub-populations (i.e. each density profile is divided by its progenitor's mass).  Figure \ref{density_normal} 
shows the density profile of GC1 and GC2 in green and in magenta respectively (where $M_{GC1} \geqslant M_{GC2}$).

We plot the profiles of nine selected runs. The profiles are at time $\sim{}550$ Myr since the beginning of the simulation. We see that the final density profiles {\change of the merger remnants are consistent with a single King model profile, although the two populations have different densities in the central regions. Depending on the run, we note that at small radii the {\change normalised} density of GC1 members  can be higher than that of GC2 members or {\it viceversa}. This suggests that the initial mass and density ratios affect the {\it relative} central density of the two populations in the final merger remnant} (Figure \ref{density_normal}). 
Despite the normalisation to the progenitor's mass, {\change at large radii} one curve is below the other in several panels. For example, in several plots of Figure \ref{density_normal} the magenta curve is  below the green one (see especially the bottom right panel: since the profiles are normalised to the mass of the progenitors, this is a clear signature of mass loss during the merging process).

\begin{figure*}
\centering
\includegraphics[width=.49\textwidth]{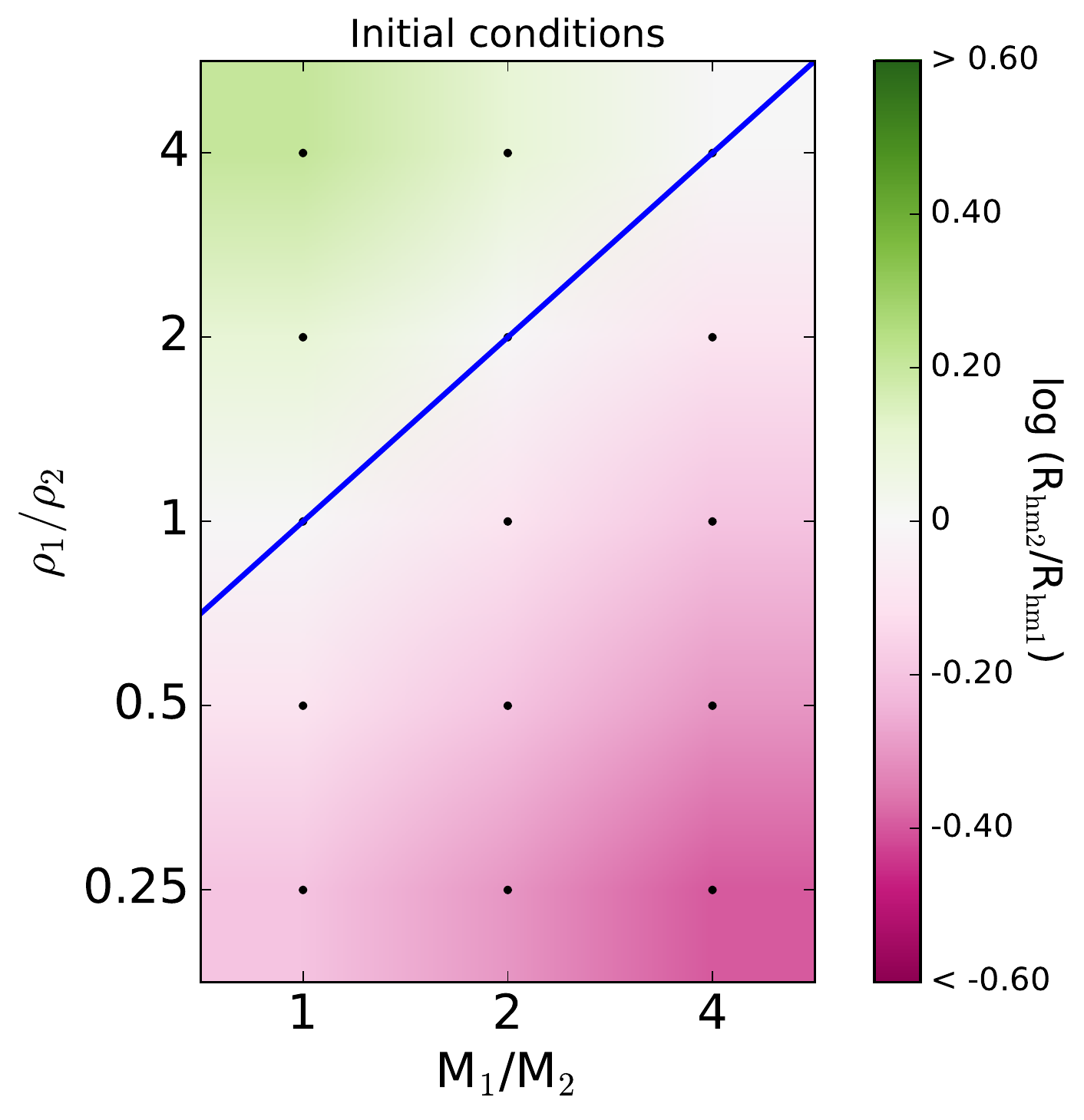}\quad
\includegraphics[width=.49\textwidth]{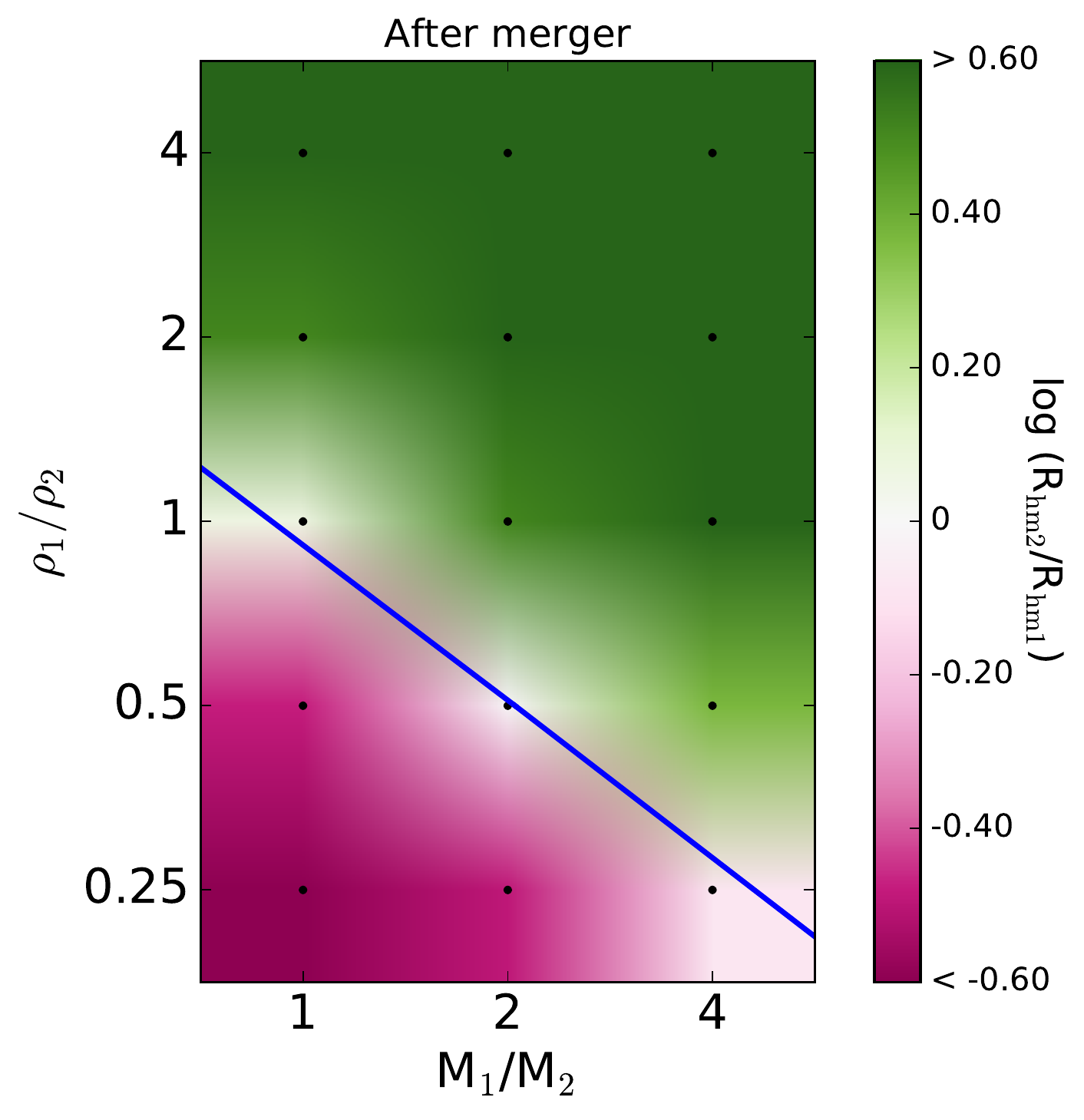}
\caption{Colour map of initial (left) and final (right) {\change ratio between the half-mass radius of GC2 and GC1 ($\log{(R_{\rm hm2}/R_{\rm hm1})}$)}.  The $x-$ and $y-$axis are the initial mass ratio and the initial density ratio of the two progenitors. The colour map quantifies the {\change relative concentration} of the two populations (in logarithmic scale), meaning the ratio of the two half-mass radii i.e. $R_{\rm hm2}/R_{\rm hm1}$. If the logarithm of {\change this value} is negative (magenta) GC2 is more centrally concentrated in the merger product; if it is positive (green), GC1 is more centrally concentrated. The blue line marks the boundary between where GC1 is more centrally concentrated (green) and the situations where GC2 is more centrally concentrated (magenta). In both plots, the actual data grid at which $R_{\rm hm2}/R_{\rm hm1}$ is evaluated is marked with black dots, the colour map is then smoothed  via interpolation in order to highlight the general trend. Note that the $x-$ and $y-$axis are effectively logarithmic.}
\label{keyplot}
\end{figure*}

Figure \ref{keyplot} is a {\change colour map of the relative concentration of the two progenitors, defined as `$\log{(R_{\rm hm2}/R_{\rm hm1})}$', i.e. the logarithm of the ratio between the half-mass radius of GC2 and GC1, in the initial conditions and at the end of the simulations, for the whole grid of runs.}
The plot on the left-hand side in Figure \ref{keyplot} shows the {\change ratio between the half-mass radius of GC2 and GC1} in the ICs, the plot on the right-hand side shows the {\change same quantity} after the merger. 
From the right panel in Figure \ref{keyplot}, we see that when the initial densities are equal, the more massive progenitor {\change dominates the central part of the merger remnant} and the less massive progenitor is more extended in the merger remnant.  
If the progenitors have equal masses, the denser progenitor is more concentrated in the remnant. 
To compensate for an unequal mass ratio, the less massive progenitor must have a density larger by roughly the factor by which its mass is lower. 
If the {\change smaller mass} progenitor is $1/A$ as massive, its initial density must be $A$ times greater or alternatively, its radius must be $A^{-2/3}$ as large as the more massive one.

\begin{figure}
\centering
\includegraphics[width=.49\textwidth]{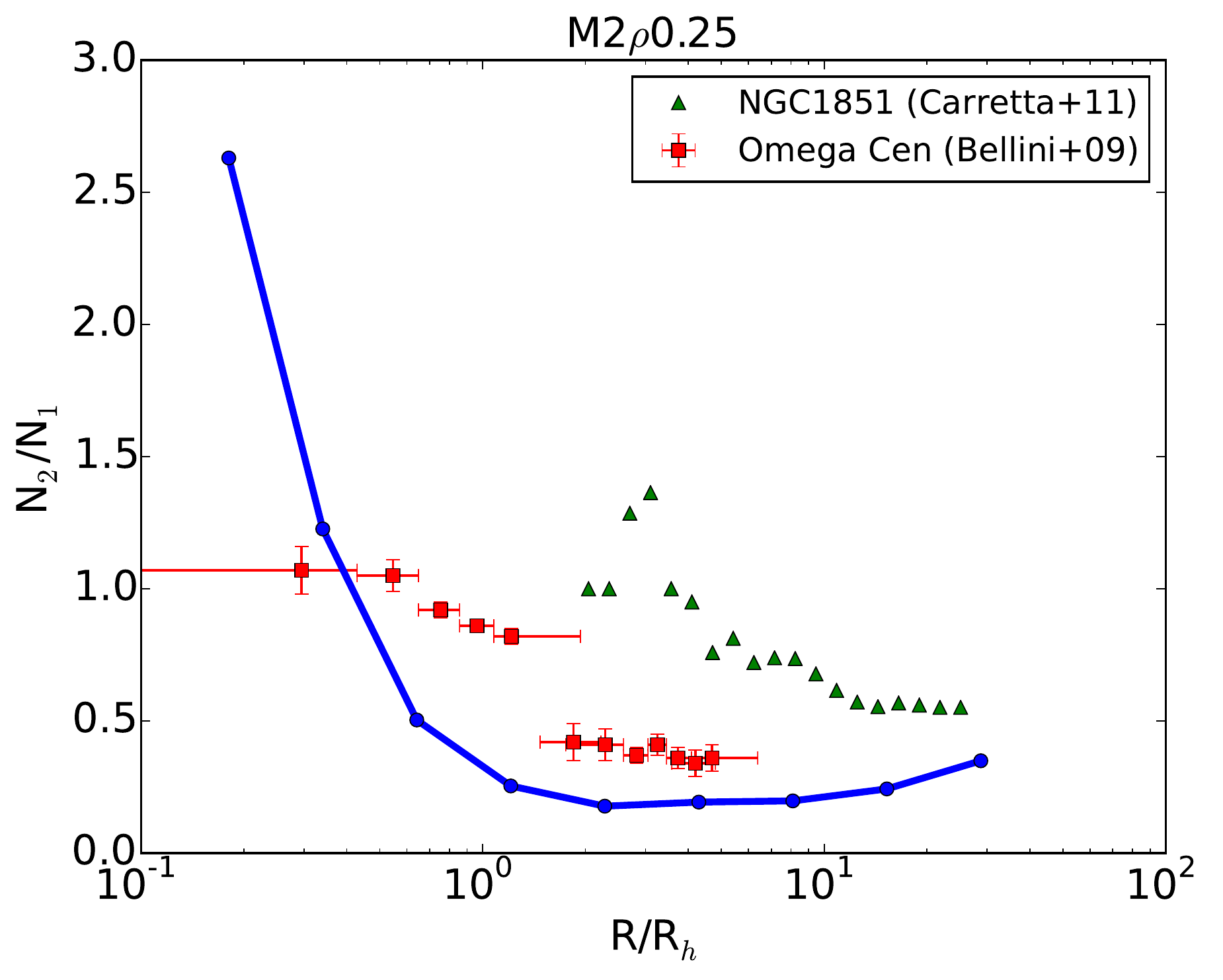}\quad
\includegraphics[width=.49\textwidth]{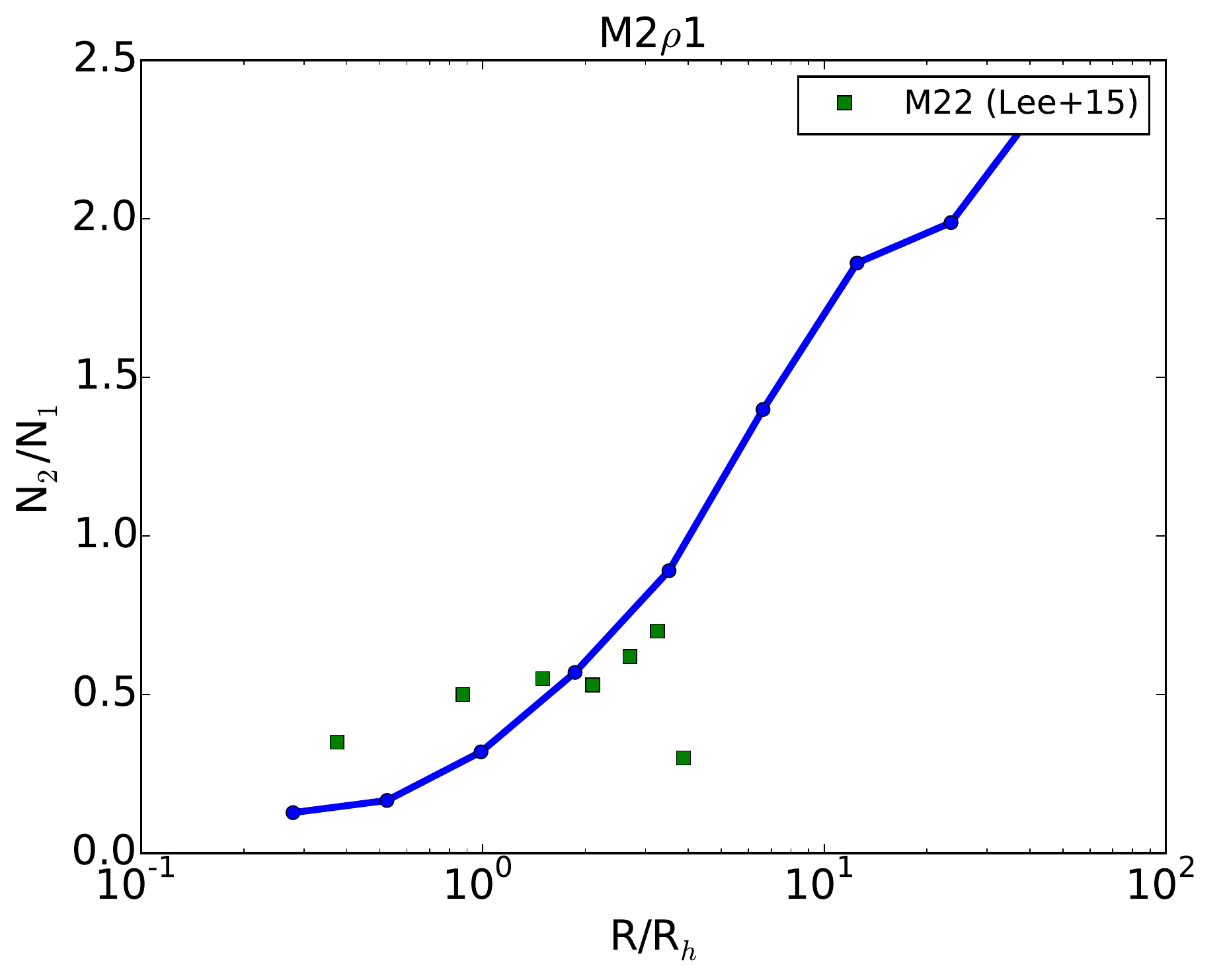}
\caption{Ratio of the minority ($N_2$) to the majority ($N_1$) population versus the radial distance from the centre. The blue solid line indicates our simulated models M2$\rho{}$0.25 (top panel) and M2$\rho{}$1 (bottom panel). The data points refer to observations (\citealt{Bellini09} for $\omega$ Cen, \citealt{Carretta11} for NGC~1851, and \citealt{Lee15} for M~22).  $N_2/N_1$ is normalised to the half-mass radius and to the half-light radius for the simulations and for the observations, respectively.}
\label{obs}
\end{figure}

In Figure~\ref{obs} we compare the number ratio of sub-populations ($N_2/N_1$) in our simulated GCs  with the observations. Specifically, we plot the ratio of the minority ($N_2$) to the majority ($N_1$) population against the radial distance from the centre, normalised to the half-mass (or half-light) radius.   
Observational data of three GCs are compared with our simulations: in $\omega$ Cen the metal-rich population is the most centrally concentrated and is the minority population \citep{Bellini09}, in NGC 1851 \citep{Carretta11} and M22 \citep{Carretta11} the metal-poor population is the more centrally concentrated (note that crowding prevents observing the very central regions of NGC 1851).  In M22 the metal-rich population is the minority, while in NGC 1851 the metal-poor population is the minority. 

The two runs shown in Figure~\ref{obs} (M2$\rho{}$0.25 and M2$\rho{}$1) were not tuned to reproduce the observations, but  follow the same trend as the data. 
In our simulations, the metallicity is only a tag:  in the top panel of Figure~\ref{obs} we use the same model (with a denser minority population) to match cases where the minority population is more concentrated, but the minority population is metal rich in $\omega$ Cen and metal poor in NGC~1851.  We adopt a different progenitors model (with equal density) for M~22, where the minority population (metal-rich) is less concentrated.

\subsection{Rotation}\label{results2}

Rotation is observed in nearby GCs \citep{Anderson05, vandeBosch06, Bellazzini12, Lardo15, Fabricius14}, {\change which can} arise from a variety of mechanisms \citep{BertinVarri08, VarriBertin12, Bianchini13, vesperini14}. 
{\change While } there is no connection demonstrated between rotation and multiple populations, \cite{AmaroSeoane13} pointed out that $\omega$~Cen, M~22, and NGC~2419
are among the most flattened clusters.

{\change Flattening has been detected in several galactic GCs \citep{WhiteShawl87,ChenChen10} and could be explained by several physical factors, such as pressure anisotropy or external tides \citep{VanDenBergh08}. Another possible justification for the flattening is the internal rotation of GCs \citep{Fabricius14}.} A correlation between flattening and iron-complex multiple populations would favour the merger scenario.   

In this section, we look at the detailed kinematics of our merger remnants, as a function of mass and density ratios. {\change We want to quantify their amount of rotation and see whether their degree of flattening correlates with rotation.}

\begin{figure*}
\centering
\includegraphics[width=.3\textwidth]{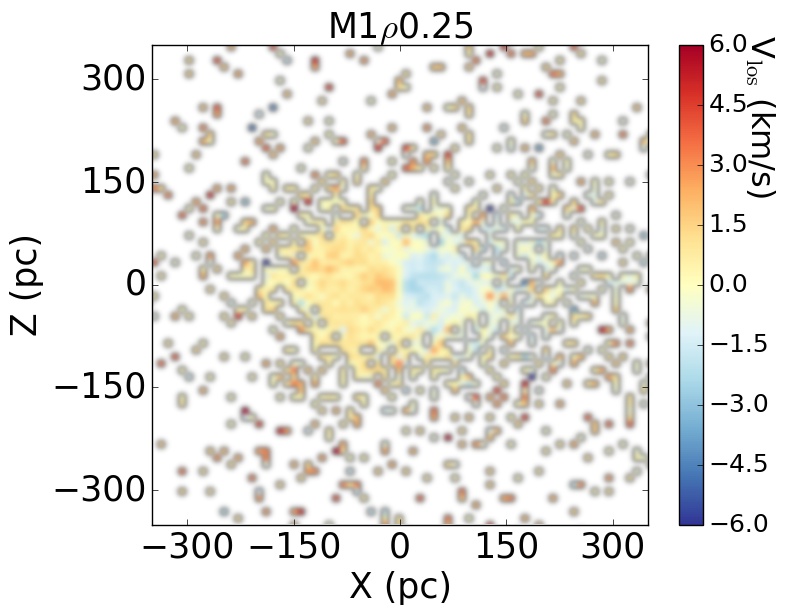}\quad
\includegraphics[width=.3\textwidth]{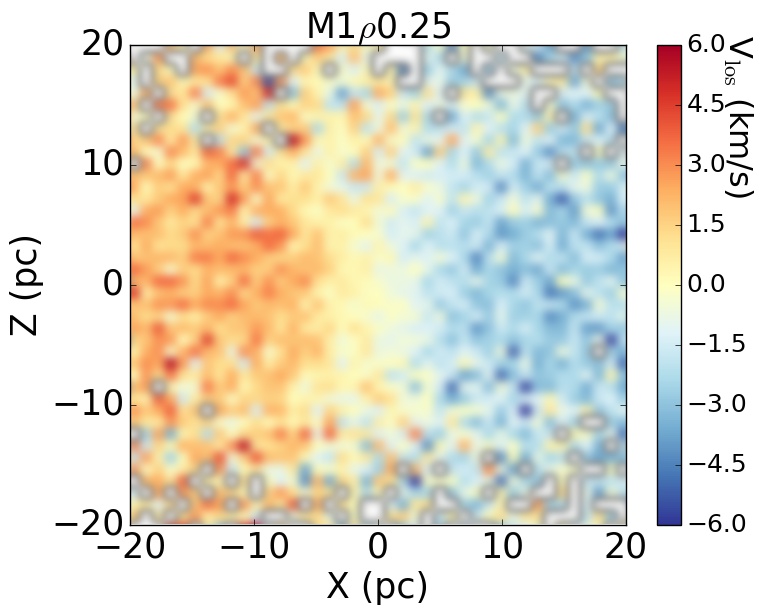}\quad
\includegraphics[width=.3\textwidth]{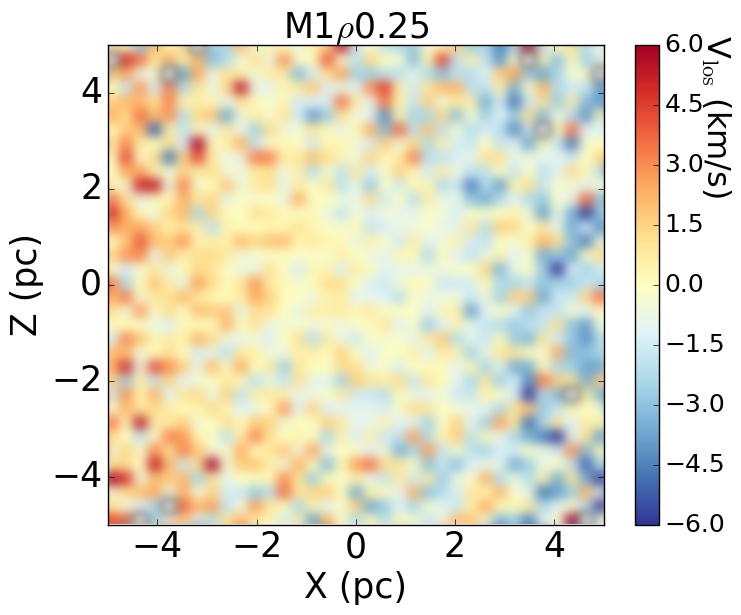}
\medskip
\includegraphics[width=.3\textwidth]{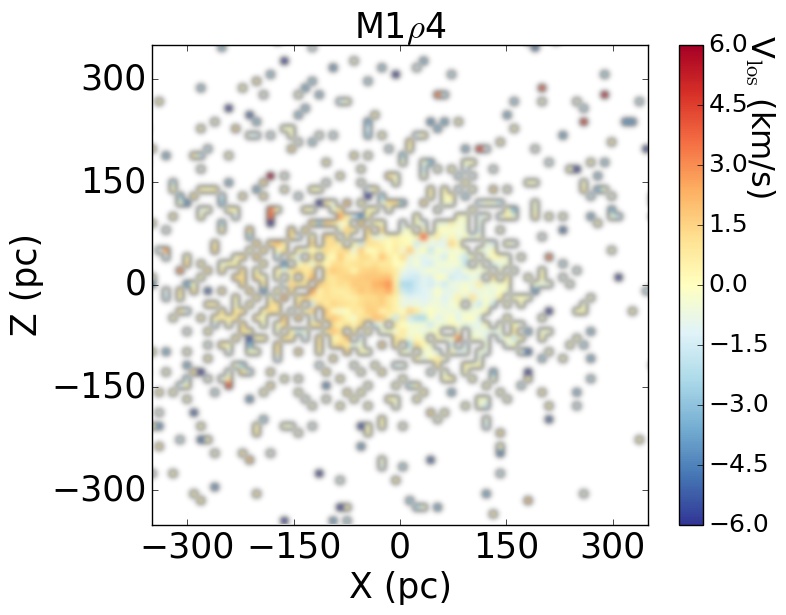}\quad
\includegraphics[width=.3\textwidth]{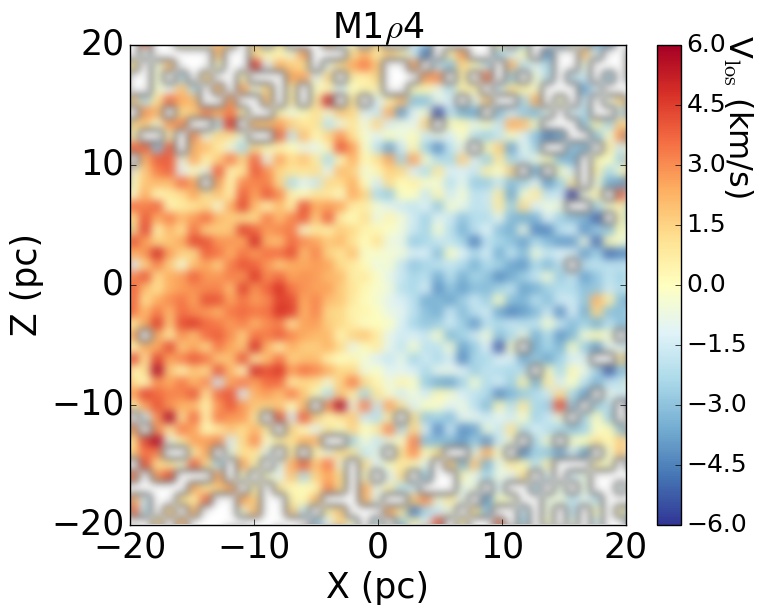}\quad
\includegraphics[width=.3\textwidth]{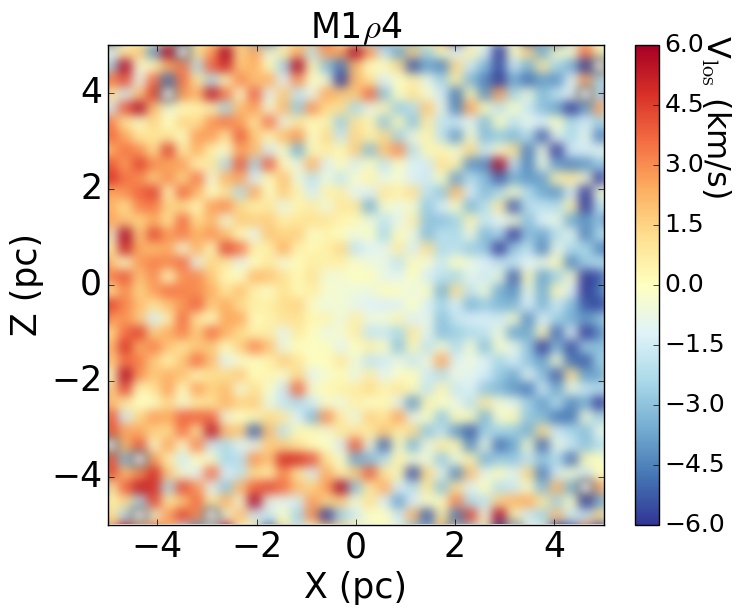}
\caption{Line-of-sight velocity maps at different scales at t= 550 $\myr$ for the case with equal mass ratio between the progenitors and $\rho_1/\rho_2$=0.25 (top row),  $\rho_1/\rho_2$=4 (bottom row). From left to right, we zoom in the central parts of the remnant.
The largest scales (left-hand columns) show a clear flattening.  
Examining these colour maps, 
the rotation within 5 pc is generally solid body (colour is changing), it becomes differential at 5-10 pc (the colour stays constant
outside this radius in the rotation plane) and it is {\change cylindrical} everywhere (weak
or no colour trend vertically).  The similarity of all the maps reveals that these are common features of mergers. }

\label{velmap500_1}
\includegraphics[width=.3\textwidth]{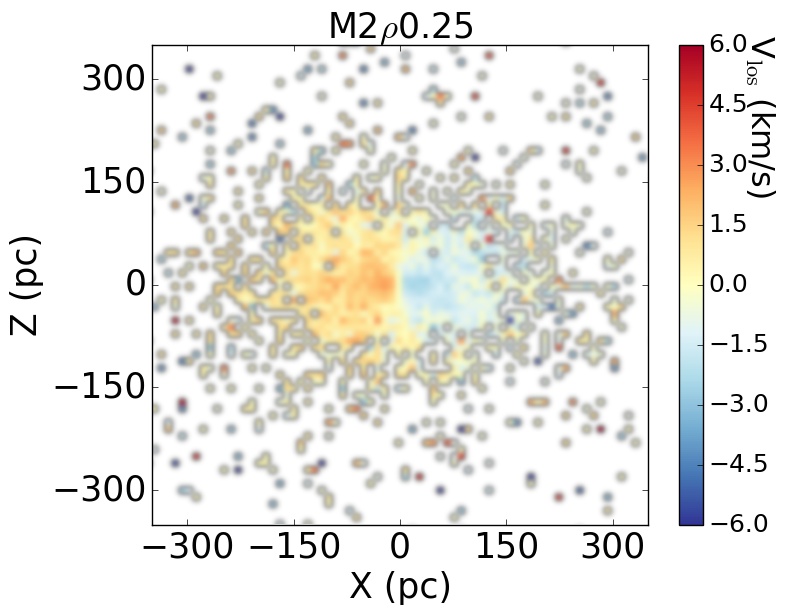}\quad
\includegraphics[width=.3\textwidth]{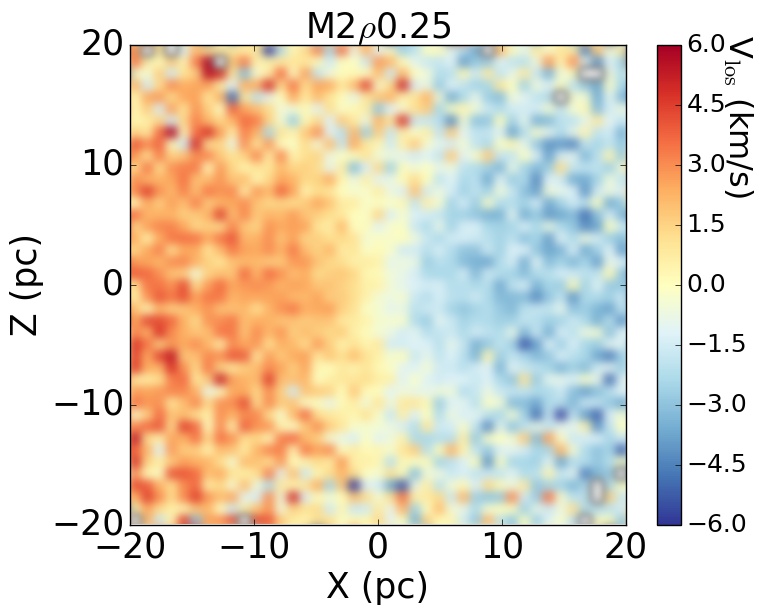}\quad
\includegraphics[width=.3\textwidth]{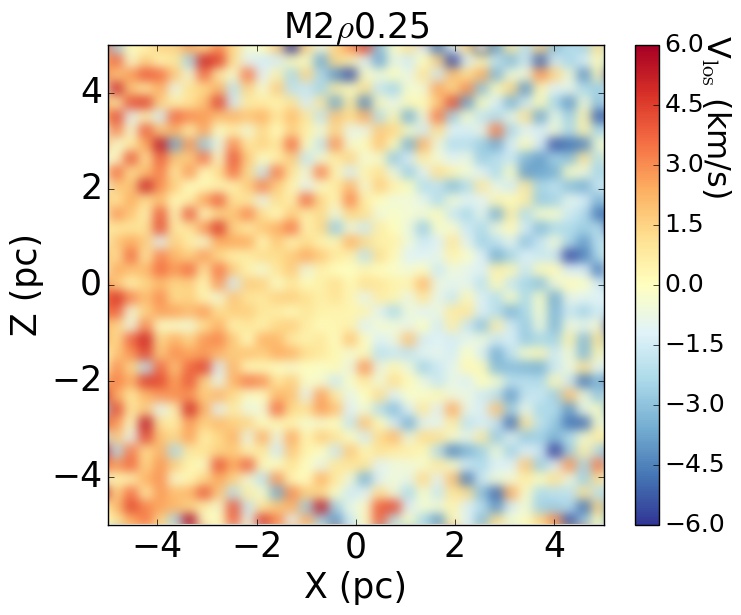}

\medskip
\includegraphics[width=.3\textwidth]{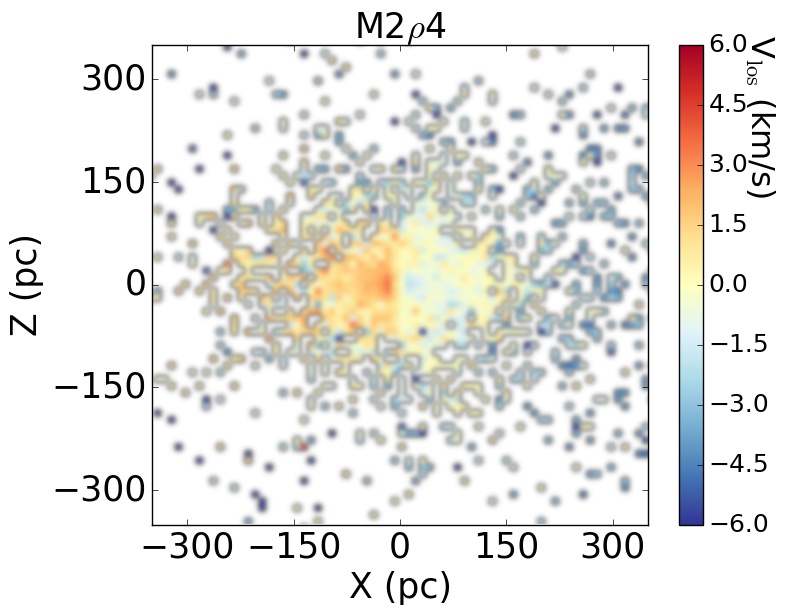}\quad
\includegraphics[width=.3\textwidth]{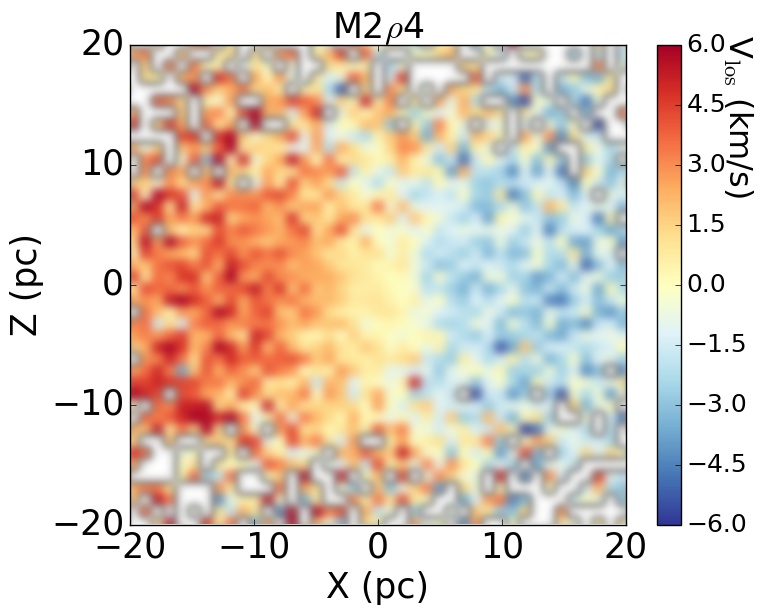}\quad
\includegraphics[width=.3\textwidth]{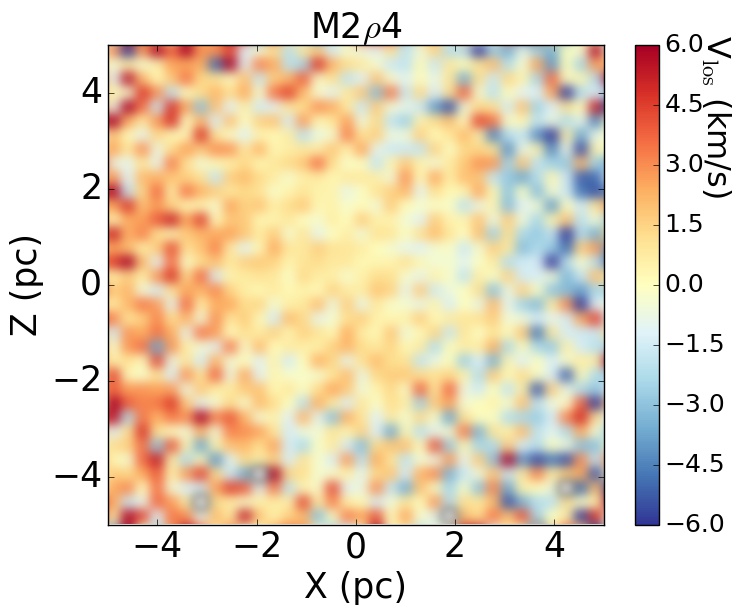}
\caption{The same as Figure~\ref{velmap500_1}, but for the case with 2:1 mass ratio between the progenitors and $\rho_1/\rho_2$=0.25 (top row),  $\rho_1/\rho_2$=4 (bottom row).}

\label{velmap500_2}

\end{figure*}

\begin{figure*}
\centering
\includegraphics[width=.3\textwidth]{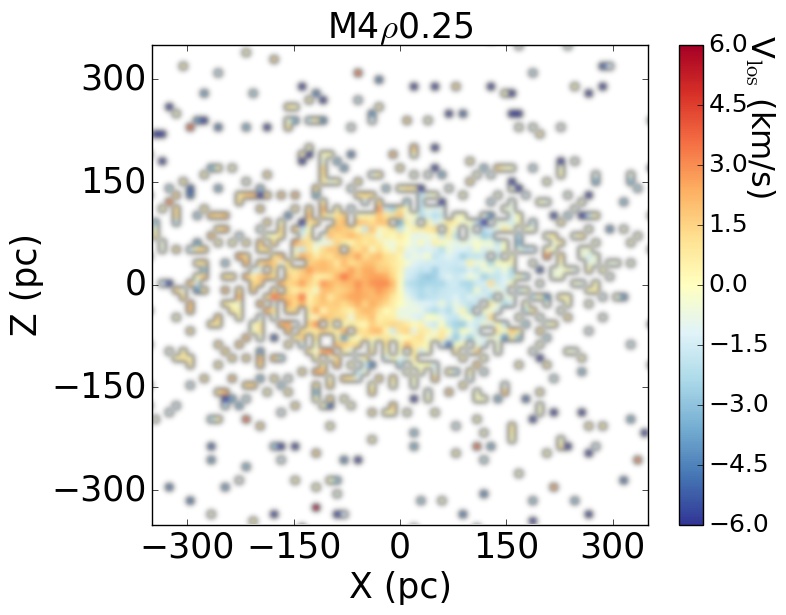}\quad
\includegraphics[width=.3\textwidth]{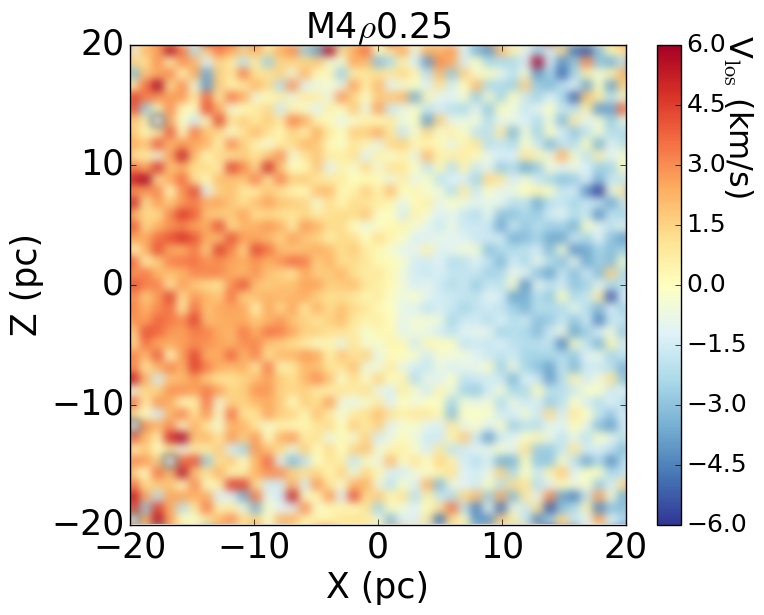}\quad
\includegraphics[width=.3\textwidth]{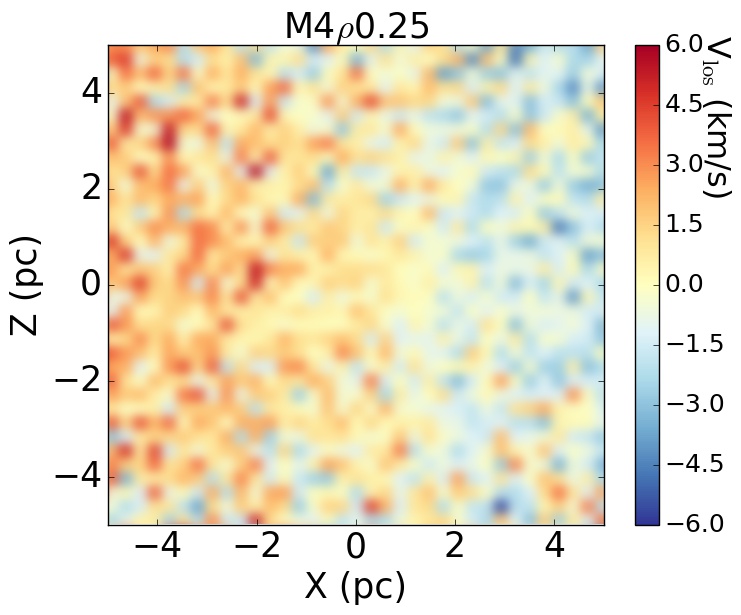}

\medskip
\includegraphics[width=.3\textwidth]{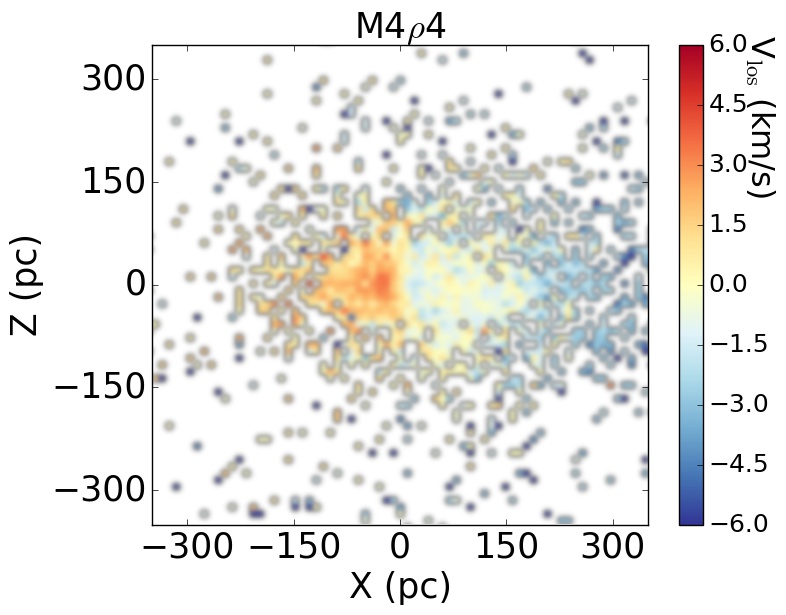}\quad
\includegraphics[width=.3\textwidth]{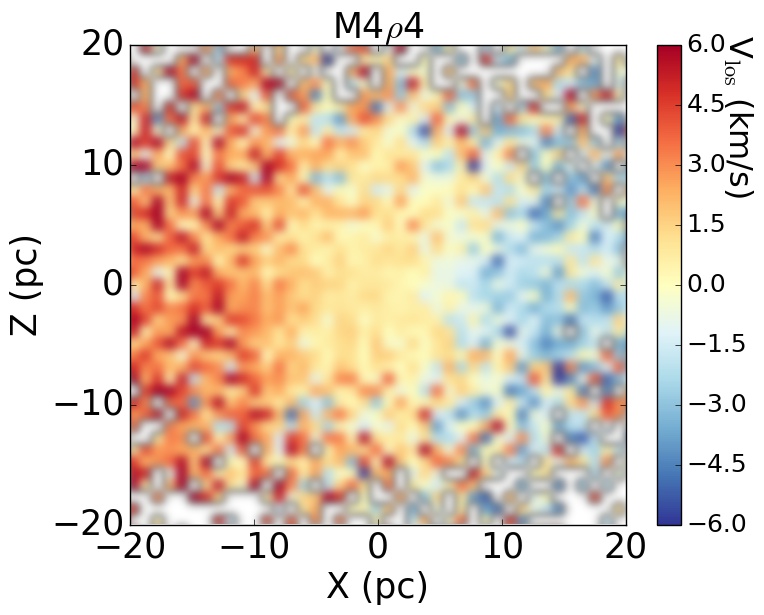}\quad
\includegraphics[width=.3\textwidth]{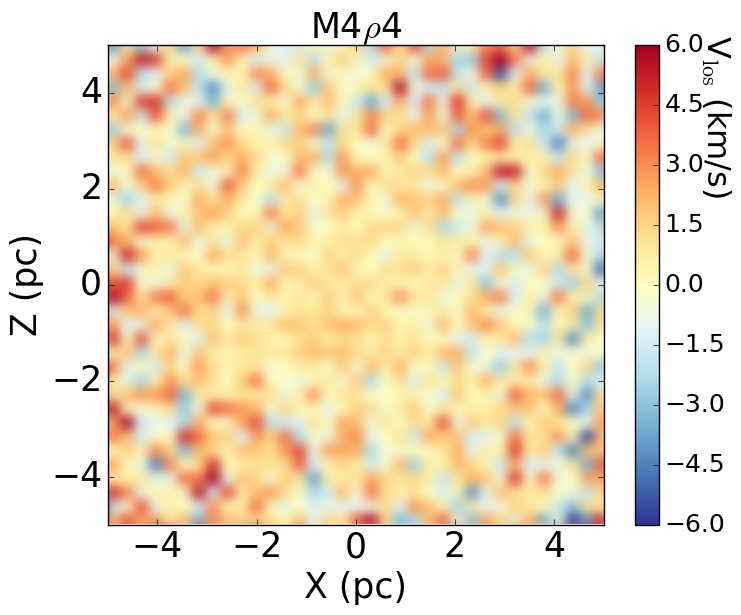}
\caption{The same as Figure~\ref{velmap500_1}, but for the case with 4:1 mass ratio between the progenitors and $\rho_1/\rho_2$=0.25 (top row),  $\rho_1/\rho_2$=4 (bottom row).}
\label{velmap500_4}

\end{figure*}

{\change All of our merger remnants have rotation, as a consequence of the parabolic orbits of their progenitors.} 
In Figs. \ref{velmap500_1}, \ref{velmap500_2} and \ref {velmap500_4}, we show velocity maps for the complete range of initial mass ratios and the
{\change limiting} cases of density ratios $\rho{}_1/\rho{}_2=0.25,\,{}4$.  In all cases, we plot line-of-sight velocities for an observer sitting on the mid-plane perpendicular to the rotation axis. 
For comparison with the observations \citep{Fabricius14}, we used a Gaussian filter to create the velocity map, progressively zoomed from left to right. 
{\changetwo Even the largest {\change spatial} scales of the final merger state (left-hand columns) show a clear flattening and {\change the two populations have similar properties} in configuration and velocity space. }

These maps illustrate some important trends: 
the rotation within 5 pc is generally solid body, it becomes differential at $5-10$ pc, and the rotation is {\change cylindrical} everywhere. The similarity of the maps shows that these features are common to all our simulations. 
Solid-body rotation is the most probable distribution function (maximal entropy) for the relaxed core of a rotating N-body system \citep{LyndenBell67,LightmanShapiro78}. 
Observations of clusters also show solid body rotation over most of the half-mass radius and differential rotation outside  \citep{MeylanMayor86, Bianchini13, Fabricius14}.

%%%%%%%%%%%%%%%%%%%%%%%%%% VEL MAP AT 500 Myr%%%%%%%%%%%%%%%%%%%%%%%

\begin{figure}
\centering
\includegraphics[width=0.70\columnwidth]{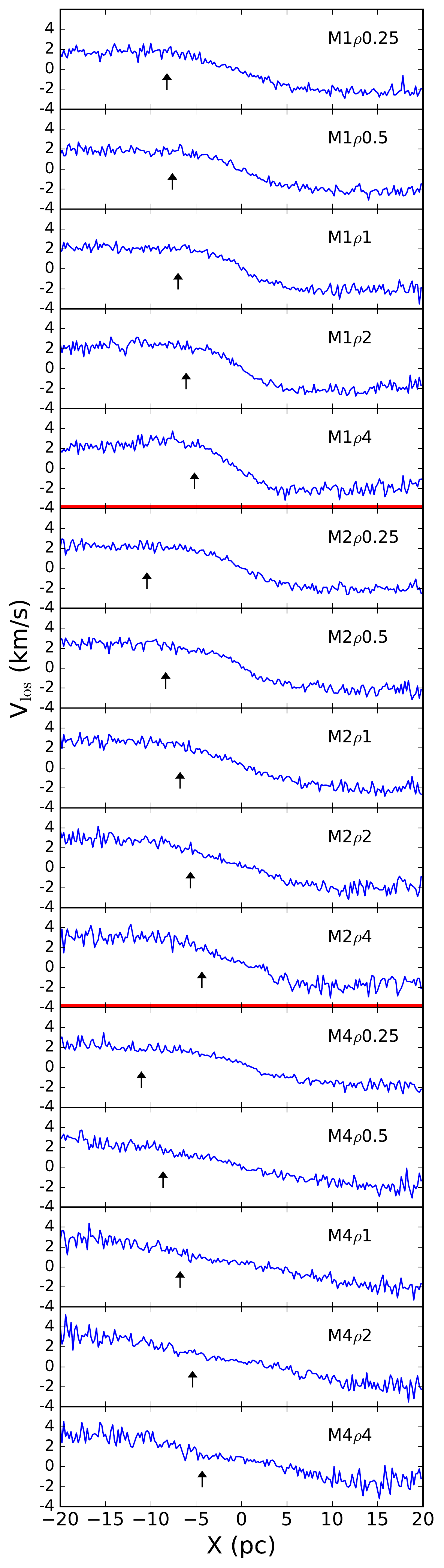}
\caption{Line-of-sight velocity profile for all the simulations for an observer in the mid plane ($V_{\rm los}$).   
The half-mass radius is denoted by an arrow in each plot. The thicker red horizontal lines divide the panels by mass-ratio (the top group has $M_1/M_2 = 1$, the central $M_1/M_2 = 2$, the bottom one $M_1/M_2 = 4$). Every group has plots for the 5 density ratios considered.}\label{rotation_curves}
\end{figure}

Figure \ref{rotation_curves} shows the line-of-sight rotation profiles of all the simulations for an observer sitting on the rotation plane. 
As in the velocity maps,  we see that the inner rotation is  
solid-body, then it becomes differential at 5-10 pc. The solid-body rotation region is more extended in the runs with higher mass ratio;  the angular momentum of the {\change less massive} object is preferentially deposited in the outskirts of the remnant.

\begin{figure}
\centering
\includegraphics[width=1.0\columnwidth]{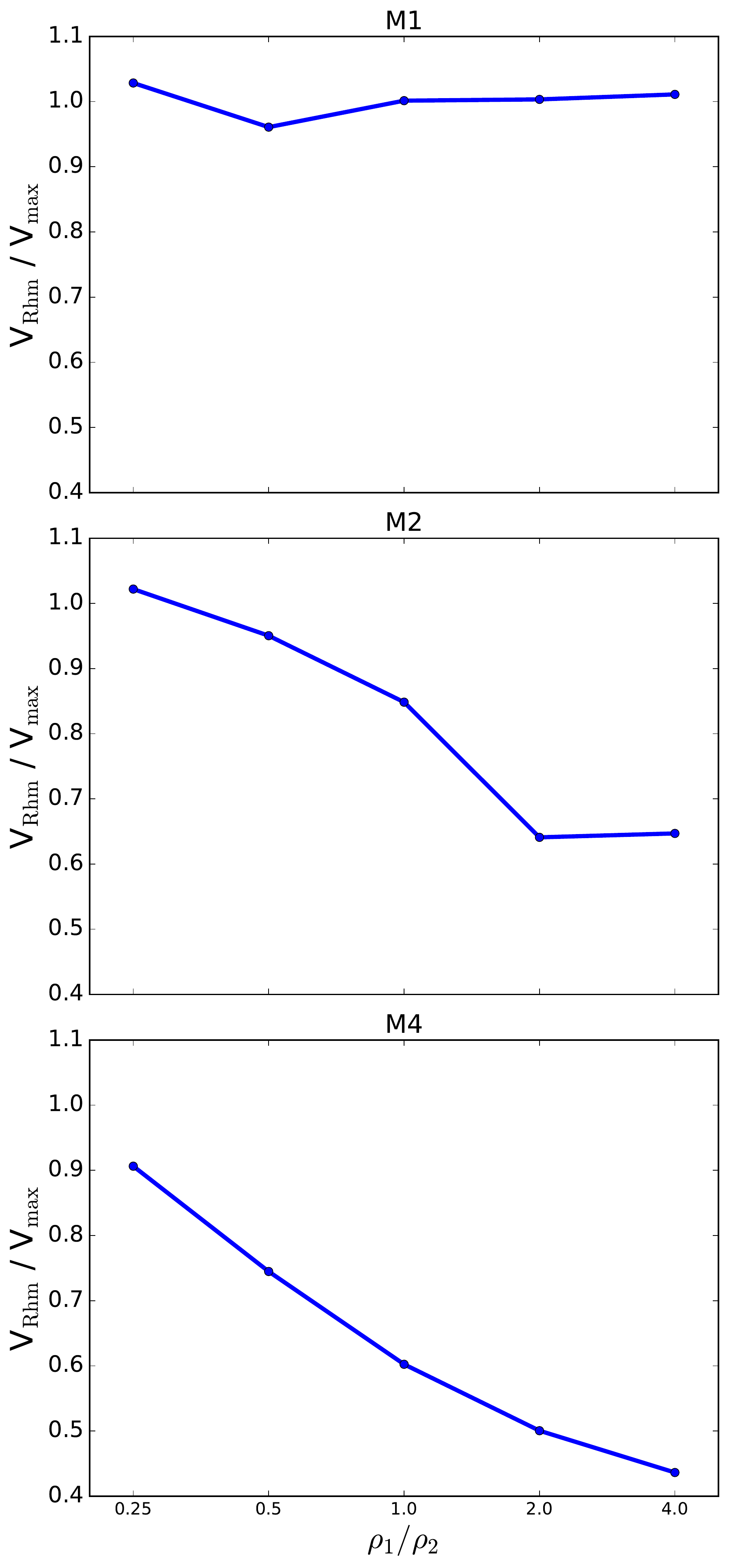}
\caption{\change Ratio of the rotation velocity at the half-mass radius to the maximum rotation velocity, $V_{\rm Rhm}/V_{\rm max}$, as a function of density ratio. From top to bottom: each panel refers to GCs with mass ratio $M_1/M_2=1,$ 2 and 4. 
Note that the $x-$axis is effectively logarithmic.}\label{3panel}
\end{figure}

{\change At the half mass radius, the merger remnant exhibits solid-body or differential rotation depending on the initial density ratio between the progenitors. In Figure~\ref{3panel}, }  
we examine the ratio of the rotation 
velocity at the half-mass radius to the maximum rotation velocity $V_{\rm Rhm}/V_{\rm max}$, as a function of density ratio.   For equal-mass ratios the quantity $V_{\rm Rhm}/V_{\rm max}$ is almost constant with respect to the density ratio (top panel of Figure~\ref{3panel}). {\change Therefore each of these model clusters have transitioned from  solid-body to differential rotation by the half mass radius.} In contrast, the {\change trend} for unequal mass ratios {\change provides} an interesting test of the model.      
{\change When the  less massive progenitor is less dense, it deposits its angular momentum in the outer parts.  In contrast, small-mass progenitors with larger density burrow into the centre.} 
When the minority population is more concentrated, the rotation
curve will peak at roughly the half-mass radius, whereas when the minority population is less concentrated, the peak occurs outside the half-mass radius. 
{\change Therefore, for unequal-mass ratios, $V_{\rm Rhm}/V_{\rm max}$ decreases for increasing values of the density ratio. }

\begin{figure}
\centering
\includegraphics[width=\columnwidth]{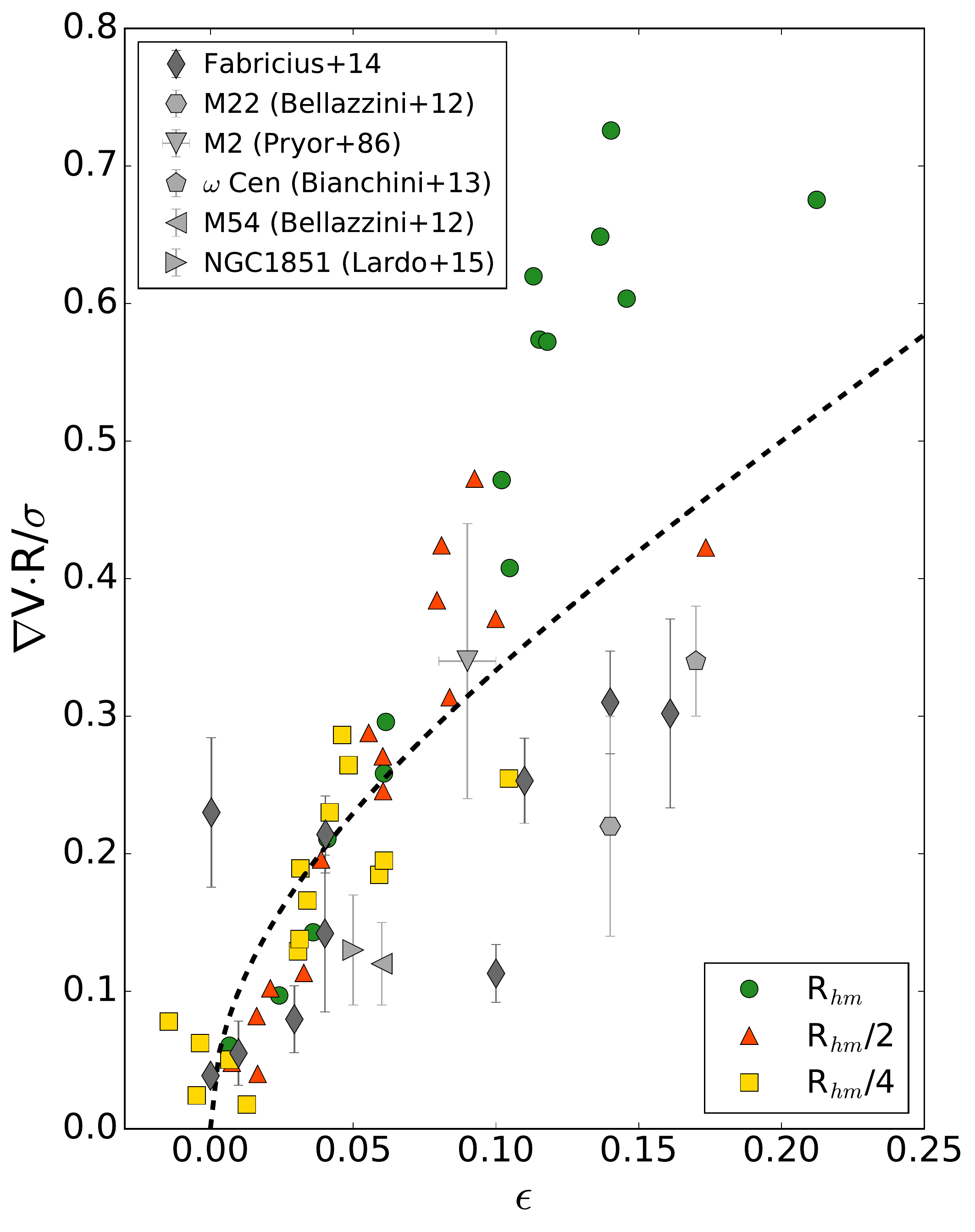}
\caption{($ \nabla V \cdot R /\sigma$, $\epsilon$) for each model at different values of $R$. {\change The observational points (grey) are indicated in legend in the top-left corner}.  The black dashed line indicates the behaviour of an isotropic oblate rotator.}
\label{gradVr_sigma}
\end{figure}

{\change In order to compare the outcomes of our simulations with observations, we study now the ($V/\sigma$, $\epsilon$) diagram, which relates the ratio of the rotation velocity $V$ and random motion $\sigma$ to the ellipticity $\epsilon$, which measures the flattening.

The expectation for isotropic rotators are derived from the tensor virial theorem \citep{Chandra69}.  The rotation velocity is the square root of the mass weighted streaming velocity squared.  The velocity dispersion is the unordered kinetic energy.  If the mass is stratified on concentric similar ellipsoids, the density profile drops out \citep{Roberts62,ChandraLebovitz62} and the ratio of the ordered kinetic energy to the unordered one (or its square root $V/\sigma$) is a function only of the ellipticity $\epsilon$ \citep{Binney78}.   The application to elliptical galaxies is straightforward  since $V$, $\sigma$ and $\epsilon$ are all nearly constant with radius \citep{Sauron}.   

For GCs,  $\epsilon$ has a greater variation  with radius and  $V$ is rising with an asymptote at a radius beyond the observations. 
Hence, we look at how well  `proxy' and `measured' rotations relate to one another in the simulated merger remnants.  As always with proxies
and dimensionless parameters that vary with radius, the
results will be mixed.

In Figure \ref{gradVr_sigma} we plot the ($V/\sigma$, $\epsilon$) diagram, following the prescription of \cite{Fabricius14} as proxies for $V$ and $\epsilon${\changetwo , including both data from our simulations and observed GCs.}
\cite{Fabricius14} fit  a plane $V(x,y) = ax +by + V_{\rm sys}$ (where $V_{\rm sys}$ is the systemic velocity) to the velocity fields  to determine the 
central velocity gradient, $|| \nabla V || = \sqrt{a^2+b^2}$.  They take velocity dispersions $\sigma$ and half-light radii $R_{hl}$ from \cite{Harris96} to create a proxy for rotational velocity $ \nabla V \cdot R_{hl}$, and find $V/\sigma$ increasing with ellipticity. 
In our case, we define $V$ in similar way ($ \nabla V \cdot R$) leaving though $R$ as free parameter, with the intent to explore how this proxy for $V$ depends on the radius used to define it. $\nabla V$ is also calculated within the radius considered each time. Specifically, in Figure \ref{gradVr_sigma} we considered three different values of radius $R$, that are $R_{\rm hm}$, $R_{\rm hm}/2$, and $R_{\rm hm}/4$. Our choice is justified by Figure \ref{rotation_curves}, where the solid-body rotation shifts to differential rotation at radii varying from  $\sim{}0.5 \, R_{\rm hm}$ to $\sim{}1.5 \, R_{\rm hm}$.

$\sigma$ in our case is just the line-of-sight  velocity dispersion. As for the ellipticity, we  follow the prescriptions found in \cite{Fabricius14} and calculate ellipticity values ($\epsilon$) from the eigenvalues ($\lambda_1$, $\lambda_2$) of the covariance matrix of stellar positions (within the relevant radial dimension), i.e. $\epsilon = \sqrt{1-\lambda_2/\lambda_1}$.}

Figure \ref{gradVr_sigma} shows that the result strongly depends on the choice of radius. The  $V/\sigma$ ratio increases with ellipticity, but ellipticities  and $V/\sigma$ both increase with radius.  If we looked at Figure~\ref{rotation_curves}, we might guess that something close to $\nabla V \cdot R_{\rm hm}$ would be the best proxy and certainly would not have guessed that the plot using $ \nabla V \cdot R_{\rm hm}/4$ would look most like the oblate rotator {\change (dashed line in the plot) and would be most in agreement with the data from \cite{Fabricius14} }.  
Thus, the choice of rotational velocity in a ($V/\sigma$, $\epsilon$) diagram is not unique\footnote{{\change The proxy for $\nabla V$ adopted by \cite{Fabricius14} would always be higher than the true rotation velocity at the radius $R$, because it comes from the best linear fit to the velocities within $R$ and the second derivative of $V$ with respect to $R$ is negative (the rotation curve is flattening).  }}. 

\begin{figure}
\centering
\includegraphics[width=\columnwidth]{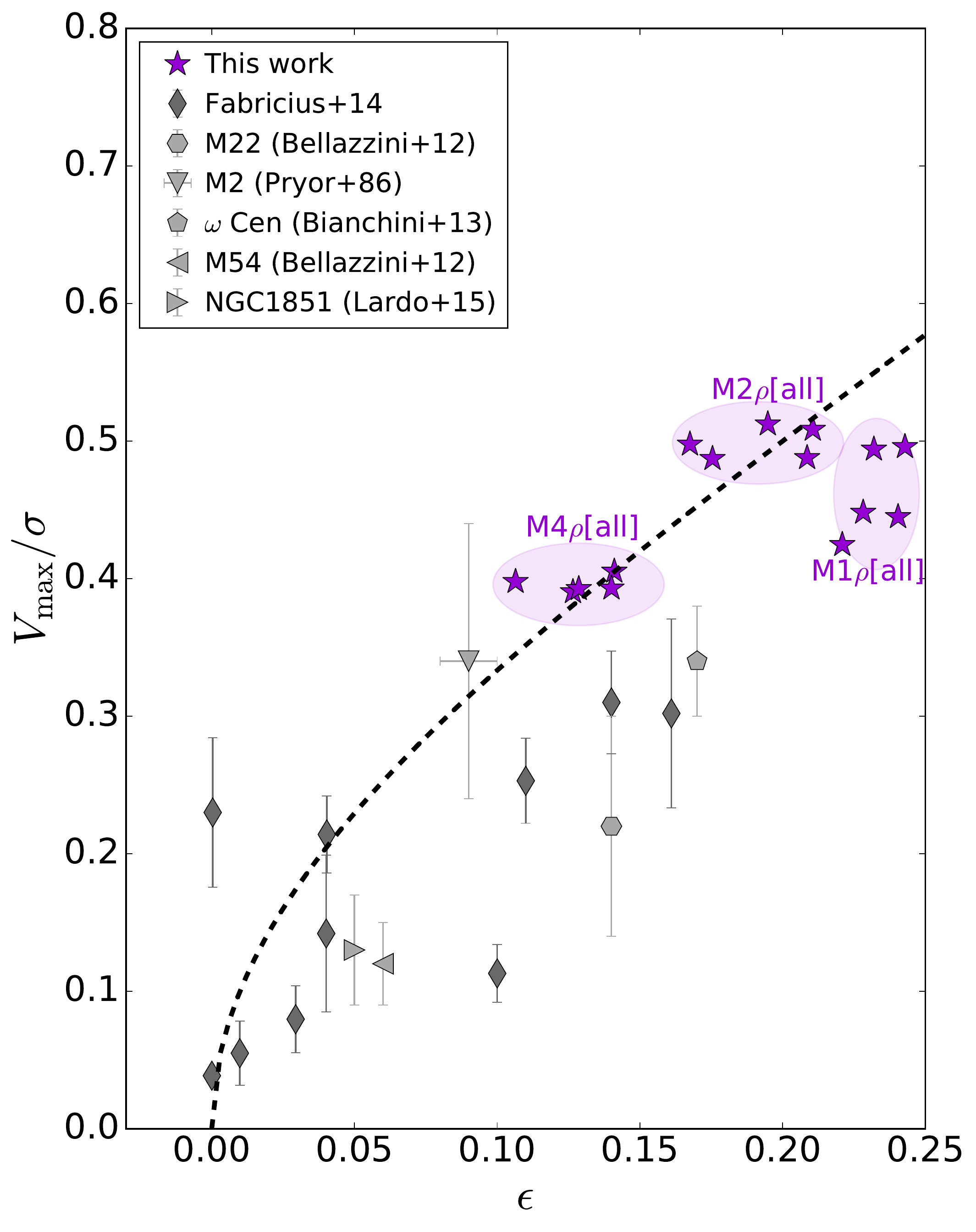}
\caption{\change ($V/\sigma$, $\epsilon$) diagram. The purple stars refer to this work, using the maximum rotation velocity ($V_{\rm max}$) for $V$.   The observational points (grey) are indicated in the legend.  The dashed black line shows the behaviour of isotropic oblate rotators.}
\label{Vmax_sigma}
\end{figure}

{\change Another possibility is to take as the rotation velocity the maximum rotational velocity.} 
In Figure~\ref{Vmax_sigma}, we plot the ellipticity $\epsilon{}$ of our simulated merger remnants versus $V_{\rm max}$/$\sigma$, where $V_{\rm max}$ is the maximum rotational velocity  (see Figure~\ref{rotation_curves}). The result (shown as star symbols in Figure \ref{Vmax_sigma}) compares favourably with the oblate rotator curve and observations. Having set the initial orbits to parabolic, the values of ($V_{\rm max}$/$\sigma$, $\epsilon{}$) for the simulated GCs are all in the same portion of the oblate rotators curve. With time, the merger remnants will radiate away angular momentum through two-body encounters \citep{FallFrenk85}. This will make them slide down on the curve to lower ellipticity and $V/\sigma$ values, closer to the observational data, {\change  because rotation and ellipticity will decrease significantly as soon as the system relaxes and the two populations mix completely (velocities will isotropise and angular momentum will diminish)}.

{\change In Figure ~\ref{gradVr_sigma} and }Figure ~\ref{Vmax_sigma}, we plot not only the observational sample of \cite{Fabricius14}, whose 11 GCs do not show any significant spread in Fe abundance, but also data of some iron-complex GCs (M~22 and M~54 from \citealt{Bellazzini12}, M~2 from \citealt{pryor86}, $\omega{}$~Cen from \citealt{Bianchini13}, NGC~1851 from \citealt{Lardo15}). {\change To derive the value of $V_{\rm max}/\sigma$ for the iron-complex GCs, we use the double mean velocity amplitude (i.e. $A_{\rm rot}$) which is considered a good representation of $V_{\rm max}$ \citep{pancino07}.}  From the kinematical point of view, the iron-complex GCs for which $V/\sigma$ and $\epsilon$ are available do not stand out in comparison with the sample of \cite{Fabricius14}.

\section{Discussion and Conclusions}\label{discussion}

In this section, we  discuss the results of our simulations in light of the observational properties of iron-complex GCs.  We focus on GCs with multimodal iron-complex abundances because they have unique tags that can be mapped to possible progenitors. 

In the merger scenario, {\change we find that} the minority population is less centrally concentrated unless the initial density of the {\change less massive} progenitor is greater by more than the mass ratio. 
In M22, the minority is metal poor and extended.  The distribution compares well to equal density progenitors with a mass ratio of 2:1.
In $\omega$ Cen and NGC1851 the less massive population is more centrally concentrated than the majority population \citep{Bellini09}. 
Merging only cares about metallicity if there is a correlation between metallicity and mass or density.  
The minority is metal rich in $\omega{}$ Cen, while it is metal poor in NGC 1851. 
In light of our results, this means these GCs can be the result of a merger only if the less massive progenitor was the denser one.  These trends are best fit when the mass ratio is 2:1 and the {\change less massive} progenitor  is four times as dense as the {\change more massive} one.

\begin{figure}
\centering
\includegraphics[width=\columnwidth]{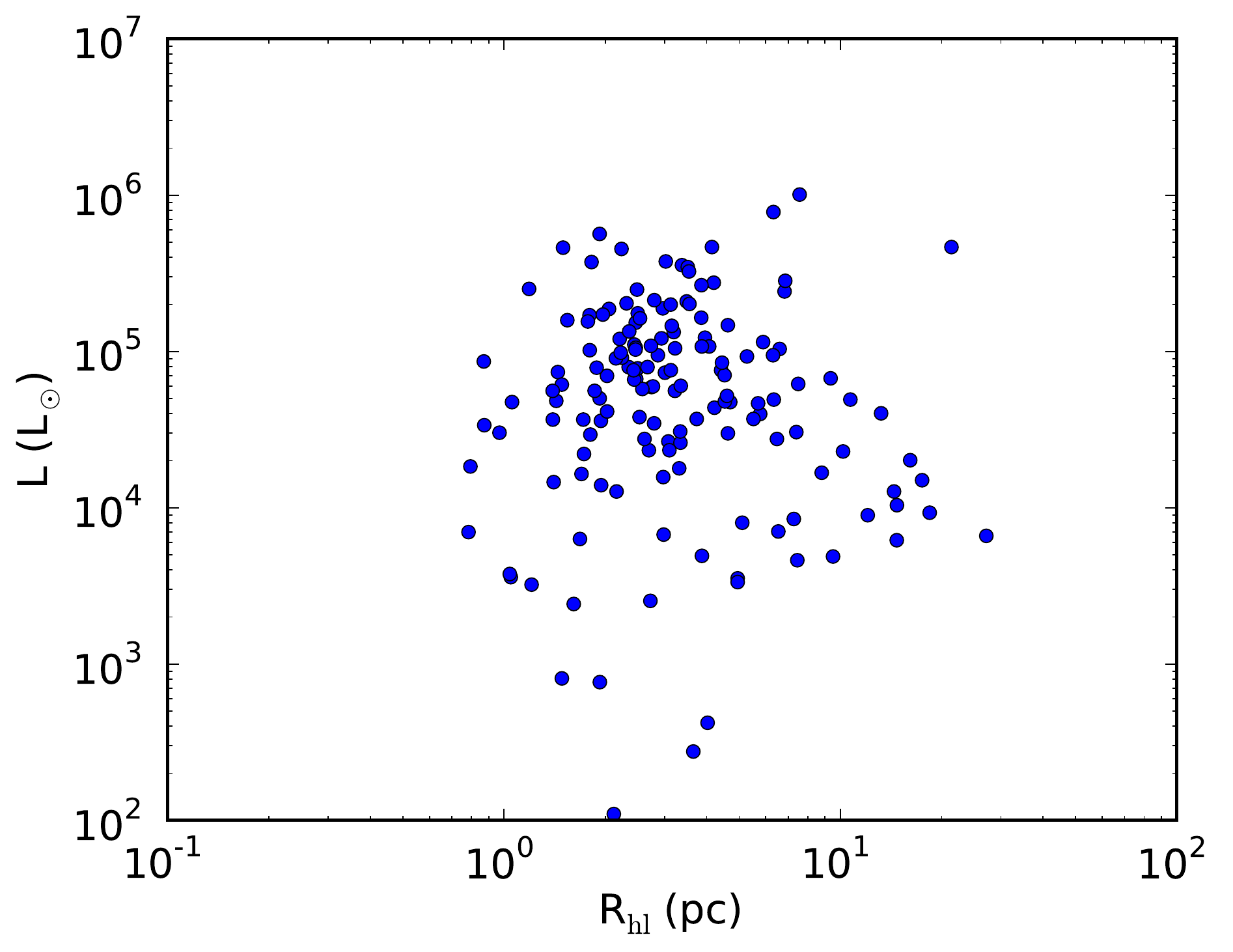}
\caption{Total luminosity versus half-light radius of Milky Way GCs from \protect \cite{Harris96}}\label{fig:fig11}.
\end{figure}

{\changetwo It would be instructive to check with observations whether the less massive progenitors are denser than more massive ones. Figure ~\ref{fig:fig11} shows the relation between luminosity (as a proxy for mass) and half-light radius in present-day GCs, from the catalogue of \cite{Harris96}. This figure shows that there is no correlation between luminosity (hence, mass) and size in present-day Milky Way GCs. From this fact, we cannot conclude very dense but small mass progenitors would be common, if  merger progenitors  were like present-day GCs. However, we also warn that considering present-day GCs as representative of merger progenitors is rather hazardous.}

The kinematical signatures of the merger remnant are similar to those observed in GCs.  In our simulated remnants: 1) the velocity dispersion is isotropic, 2) the merger product rotates close to solid body in the inner parts, then becomes differential, 3) rotation is {\change cylindrical}, 4)  at the half mass radius, the merger remnant exhibits solid-body or differential rotation depending on the initial density ratio between the progenitors, 5) the flattening of the remnant is consistent with rotation. Both  $\epsilon$ and $V$ vary over radius, so defining appropriate values for a ($V/\sigma$,$\epsilon$) plot is difficult.  Different choices move points
around in that plot, but the  correlation between flattening and rotation in the remnants is similar to the expectations from the tensor virial theorem \citep{Binney78}.

As we already anticipated in the introduction, the most severe drawback of the merger scenario is that the relative velocity between two clusters must be sufficiently low to merge. Here `sufficiently low' means that their relative velocity cannot be much larger than their velocity dispersion.
The velocity dispersion of GCs is $\approx{}3$ per cent of the velocity dispersion of stars in the field of our Galaxy. This means that GCs move too fast to merge in our present-day Galaxy.

Several studies propose that a sub-population of GCs were the nuclei of dwarf galaxies, with $\omega{}$ Cen as prototype \citep{Majewski00, Carraro00}.  If one GC were a nucleus, the inspiral of a second GC would create conditions similar to an unequal mass merger.

GCs can sink toward the centre of the host galaxy by dynamical friction. 
The dynamical friction timescale scales as the inverse of the mass of the GC. Thus, the smaller the GC, the longer it takes for it to sink to the centre by dynamical friction. For example, an object that has a mass of $\sim{}5$ per cent of the total mass of the host galaxy will spiral into the centre by dynamical friction in roughly a dynamical time \citep{BinneyTremaine}. 

The smallest dwarf galaxy in the Local Group with GCs is Fornax, with five clusters \citep{Larsen12}.  The most massive among these GCs has not yet sunk into the centre by dynamical friction \citep{Read06}.  
Thus, even Fornax failed to promote mergers or create a nucleus from its most massive GC.  

The Sagittarius dwarf galaxy is more promising \citep{GrattonCarrettaBragaglia12}.  At least five Milky Way GCs are thought to have been part of Sagittarius \citep{Law10}.   The velocity dispersion of Sagittarius is $\sim$ 20 km s$^{-1}$. Thus, parabolic encounters between GCs would be rare, but not impossible.
Sagittarius does have a nuclear cluster.  With a velocity dispersion of $\sim 20$ km s$^{-1}$, Sagittarius has a mass of $2 \times 10^8 \msun$ within one kpc, so another cluster could inspiral.  Most dwarf galaxies have likely dissolved in the old stellar halo of our Galaxy.   At $z=1$, there were roughly three times as many dwarf satellites as today \citep{Kazantzidis08}. So, there is some chance that several GCs merged within dwarf galaxies in the past. Quantifying the rate of such mergers is beyond the aims of this paper.

Finally, it is possible that two GCs merge slightly after their formation, when they are still part of the same progenitor molecular cloud. In this case, their relative velocity should be of the same order of magnitude as the turbulent motions inside the cloud (approximately few km s$^{-1}$), enabling a successful merger. There are clusters younger than 100 Myr that are believed to be ``caught in the act" of merging while they are still within the parent cloud \citep{Sabbi12}.  

{\change In summary, we confirm that the merger scenario may provide a viable explanation for multiple populations in iron-complex GCs.} Our simulations show that the relative concentration in the merger remnant betrays the initial density ratio of the progenitors. {\change Moreover, the } density ratio {\change of the progenitors} leaves a signature in the rotation curves that should be object of further observations.

\section*{Acknowledgements}
We thank the anonymous referee for their critical reading of the manuscript and for their comments, which helped us improving this work significantly. 
We thank Kim Venn, Giacomo Beccari, Eugenio Carretta and Raffaele Gratton for useful discussions. The simulations were performed with the Tasna GPUs cluster of  ZBOX4 at University Zurich, PLX and Eurora clusters at CINECA (through CINECA Award N. HP10B338N6 and HP10CZVZHA ). We acknowledge the CINECA Award N. HP10B338N6, HP10CZVZHA and the University of Zurich for the availability of high performance computing resources. EG acknowledges financial support through SNF grant and Foundation MERAC 2014 Travel grant.  MM acknowledges financial support from the Italian Ministry of Education, University and Research (MIUR) through grant FIRB 2012 RBFR12PM1F, from INAF through grant PRIN-2014-14 (Star formation and evolution in galactic nuclei) and from Foundation MERAC through grant `The physics of gas and protoplanetary discs in the Galactic centre'.

%%%%%%%%%%%%%%%%%%%% REFERENCES %%%%%%%%%%%%%%%%%%

\bibliographystyle{mnras}
\bibliography{biblio_complete_plusMM}

% Don't change these lines
\bsp	% typesetting comment
\label{lastpage}
\end{document}